\UseRawInputEncoding 
\documentclass[twocolumn]{aastex631}
\usepackage{amsmath}
\usepackage[normalem]{ulem}
\usepackage{bm}

\mathchardef\mhyphen="2D

\newcommand{\logpar}[4]{\bigg(\frac{\partial \log{#1}}{\partial \log{#2}}\bigg)_{#3, #4}}
\newcommand{\logd}[2]{\frac{d\log{#1}}{d\log{#2}}}
\newcommand{\parderiv}[4]{\bigg(\frac{\partial #1}{\partial #2}\bigg)_{#3, #4}}
\newcommand{\fullderiv}[2]{\frac{d #1}{d #2}}
\newcommand{\semipar}[4]{\bigg(\frac{\partial \log{#1}}{\partial #2}\bigg)_{#3, #4}}
\newcommand{\semiparu}[4]{\bigg(\frac{\partial #1}{\partial \log{#2}}\bigg)_{#3, #4}}

\newcommand{\semidu}[2]{\frac{d #1}{d \log{#2}}}
\newcommand{\apple}{\texttt{APPLE} }

\shorttitle{H-He EOS \& Miscibility}
\shortauthors{Tejada Arevalo et al.}

\graphicspath{{./}{figures/}}

\begin{document}

\title{Equations of State, Thermodynamics, and Miscibility Curves for Jovian Planet and Giant Exoplanet Evolutionary Models}

\author[0000-0001-6708-3427]{Roberto Tejada Arevalo}
\affiliation{Department of Astrophysical Sciences, Princeton University, 4 Ivy Ln, Princeton, NJ 08544, USA}
\author[0000-0001-8283-3425]{Yubo Su}
\affiliation{Department of Astrophysical Sciences, Princeton University, 4 Ivy Ln, Princeton, NJ 08544, USA}
\author[0000-0001-6635-5080]{Ankan Sur}
\affiliation{Department of Astrophysical Sciences, Princeton University, 4 Ivy Ln, Princeton, NJ 08544, USA}
\author[0000-0002-3099-5024]{Adam Burrows}
\affiliation{Department of Astrophysical Sciences, Princeton University, 4 Ivy Ln, Princeton, NJ 08544, USA; Institute for Advanced Study, 1 Einstein Drive, Princeton, NJ 08540}

\begin{abstract}    
The equation of state of hydrogen-helium (H-He) mixtures plays a vital role in the evolution and structure of gas giant planets and exoplanets. Recent equations of state that account for hydrogen-helium interactions, coupled with hydrogen-helium immiscibility curves, can now produce more physical evolutionary models, such as accounting for helium rain with greater fidelity than in the past. In this work, we present a set of tools for planetary evolution\footnote{All tables of thermodynamic quantities and derivatives are available at \url{https://github.com/Rob685/hhe_eos_misc}, along with a unified Python interface. Tutorials demonstrating the interface are also available in the repository.} that provides a Python interface for existing tables of useful thermodynamic quantities, state-of-the-art H-He equations of state, and pressure-dependent H-He immiscibility curves. In particular, for a collection of independent variable choices, we provide scripts to calculate the variety of thermodynamic derivatives used to model convection and energy transport. These include the chemical potential derived from the internal energy, which is a modeling necessity in the presence of composition gradients when entropy is the other primary variable. Finally, an entropy-based convection formalism is presented and fully described that highlights the physical differences between adiabatic and isentropic interior models. This centralized resource is meant to facilitate both giant planet structural and evolutionary modeling and the entry of new research groups into the field of giant planet modeling.
\end{abstract}


\section{Introduction}\label{sec:intro}

The structure and evolution of gas giant planets and brown dwarfs depend on the properties of hydrogen and helium at high pressures \citep{Hubbard1970, Hubbard1971, HDW1985, Burrows1997, Burrows2001, FortneyHubbard2003, FortneyHubbard2004, Militzer2008, Nettelmann2008, Nettelmann2012, Nettelmann2015, HubbardMilitzer2016, Mankovich2016, Pustow2016, Miguel2016, Debras2019, Mankovich2020, Debras2021, Mankovich2021, Howard2023b}, as well as their atmospheres \citep{Burrows1997,Fortney2011,Burrows2014,Chen2023}. The equation of state (EOS) most employed in the past was published by \citet[][SCvH95]{Saumon1995}. This EOS incorporated the so-called ``chemical picture" of the constituent species, obtained by minimizing the Helmholtz free energy of its constituents. This model estimated the gradual dissociation and ionization of hydrogen with increasing pressure, thus accounting for the increasing densities and even the predicted hydrogen first-order phase transition within Jupiter. Subsequent ab initio calculations by \citet[][MH13]{Militzer2013}, who included non-ideal interaction effects of hydrogen-helium mixtures and partial ionizations, were an important step forward. Models using updated EOSes by MH13 and \citet{Nettelmann2012} resulted in significant differences in the interior structures of Jupiter and Saturn \citep[see][]{Miguel2016, Militzer2013, HubbardMilitzer2016, Miguel2022}. Further complications arose when applying these EOSes to such modeling due to the phase separation of helium related to the metallization of hydrogen at high pressures; the presence of a higher-Z core; and the presence of a component of heavy elements in the gaseous mantle. The distribution of the latter is the subject of intense debate following the recent interpretations \citep{Leconte2012, Leconte2013, Helled2016, Wahl2017,Folkner2017, Vazan2018,Stevenson2020, Helled2022,Militzer2022,Howard2023b} of the \textit{Juno} \citep{Bolton2017a, Bolton2017b} gravity data. To this end, two primary areas of focus are an improved H-He EOS and a more accurate H-He phase separation curve. 

The first area, the H-He EOS, demands ab initio calculations of relevant thermodynamic quantities. The computational costs have impeded the creation of a comprehensive tabular EOS for a wide range of hydrogen-helium mixtures, temperatures, and pressures. As a result, current EOSes designed to be employed in structural and evolutionary codes, such as those of SCvH95, \citet{Nettelmann2012, Becker2014}, \citet[][CMS19]{Chabrier2019} and \citet[][MLS22]{Mazevet2022}, rely on idealized mixing laws to compute the entropy of H-He mixtures. A new class of models includes non-ideal effects of the entropy of mixing \citep{Militzer2013, Chabrier2021}. These terms have recently been estimated by \citet[][HG23]{Howard2023} based on the calculations of MH13 and CD21, where MH13 calculated a H-He mixture EOS for a helium mass fraction ($Y$) of 0.245. This resulted in estimates for the non-ideal corrections of the entropy and specific volume (density) for any helium mass fraction for the CMS19 and MLS22 EOSes. However, \citet{Howard2023} did not provide the corresponding correction for the internal energy.  

In the second area, the H-He phase separation curves, \cite{Lorenzen2009, Lorenzen2011}, \cite{Morales2013}, and \cite{Schottler2018} have employed ab initio techniques to determine the phase separation curves of hydrogen and helium mixtures. On the experimental front, \cite{Brygoo2021} conducted cutting-edge experiments on H-He phase separation and found significant disagreements with the ab initio calculations. Moreover, those ab initio calculations are not accompanied by corresponding EOSes, and the CMS19 and MLS22 EOSes are not accompanied by miscibility curves. Therefore, the predominant hybrid approach used in planetary evolution calculations, combining EOSes and phase separation curves from different studies, is inconsistent. However, the utility of this approach cannot be denied, and the ease of use of both EOSes and miscibility curves is of great importance. 

Between the different EOS and phase separation calculations by various authors, planetary modelers have a wide range of options.  In this paper, we collect many major EOS and miscibility databases required for giant planet evolutionary and structural calculations and describe their strengths and weaknesses\footnote{We omit the H-He EOS of \citet{Nettelmann2012} and \citet{Becker2014} due to the lack of entropy calculations available in their tables. An integration technique to obtain the entropy for these tables depends on choices of initial thermodynamic values, making the comparisons between EOSes uncertain and subsequent derivative calculations subject to this uncertainty.}. We invert each EOS on a dense grid and calculate all the necessary quantities and derivatives that play a role in the physics, structure, thermal evolution, and convection criteria within Jupiter and Saturn in particular, and giant exoplanets in general. Each EOS is calculated along pressure-temperature, density-temperature, and entropy-density axes that span the thermodynamic range of the interiors of gas giants. As discussed in Section~\ref{sec:convection}, a proper treatment of convection requires the calculation of various derivatives. In addition to the various EOS tables, we furnish useful tables of the hydrogen-helium immiscibility curves of \cite{Lorenzen2009, Lorenzen2011} and \cite{Schottler2018}. While these data are published and employed in some of the most recent planetary evolution calculations for Jupiter and Saturn \citep{Nettelmann2015, Pustow2016, Mankovich2016, Mankovich2020}, smooth fits to these data are not easily come by. Though demixing temperatures for a broad range of helium abundances are what authors such as \cite{Lorenzen2009, Lorenzen2011} and \cite{Schottler2018} have provided, easily used fits to these data are not easily generated. We provide such smooth fits here. In this way, our work provides a more centralized and standardized approach by collecting the most recent EOS and H-He immiscibility data in a self-consistent format. 

This paper is organized as follows. Section~\ref{sec:H_He} discusses the various hydrogen-helium (H-He) EOSes, compares their properties, and discusses their applications. In Section~\ref{sec:H_He_misc}, we provide an overview of the available miscibility calculations in the literature, compare the different demixing temperature calculations of each of the sources, compare the H-He demixing temperatures with different available H-He equations of state, and discuss what various authors have concluded concerning helium rain in Jupiter and Saturn based on these different demixing temperatures. Section~\ref{sec:tables} describes the different EOS table inversions, inversion techniques, interpolation methods, and thermodynamic table ranges. This section also describes the H-He EOS and miscibility code structure. In Section~\ref{sec:consistency}, we briefly discuss the thermodynamic consistency of the EOSes published by \cite{Saumon1995}, cite{Chabrier2021}, and \cite{Chabrier2019} \citep[with the non-ideal corrections by][]{Howard2023}). In Section~\ref{sec:convection}, we discuss the various derivatives necessary for convection and energy calculations in a planetary model. Section~\ref{sec:model_example} showcases an example evolutionary calculation using the code \apple \citep{Sur2024} and the tables presented here. Finally, in Section~\ref{sec:conclusion}, we provide a summary discussion and concluding remarks. In the appendices, we derive various of the relationships provided in the text.

\section{Hydrogen and Helium Equations of State}\label{sec:H_He}

Historically, the EOS of hydrogen and helium at high pressures has been divided into the ``physical'' picture and the ``chemical'' picture. In the physical picture, electrons and nuclei interact through Coulomb potentials, and the EOS is obtained by solving the resulting quantum many-body problem. In the chemical picture, atoms remain bound and interact through pair potentials. However, at ionization pressures and temperatures, electrons become unbound and delocalized. In the high temperature regime, electrons become unbound and the physical picture is tractable \citep[][]{Rogers1981, Rogers1984}.  However, in the low-temperature and high-pressure regime relevant to gas giant planets, the physical picture could not easily handle partial ionization. The chemical picture presents a simplified model a physical model. In the past, the chemical picture of SCvH95 was the most commonly used H-He EOS. However, with improvements in computation  rigorous physical (or ``ab initio'') models have now been published. Most notably, MH13, CMS19, \citet[][CD21]{Chabrier2021}, and MLS22 have published EOSes obtained from ab initio calculations. All of these models rely on the additive volume law (AVL), since calculations for arbitrary hydrogen-helium mixtures remain intractable. This approximation was improved by HG23, who have calculated nonideal interaction corrections based on the MH13 results for the CMS19 and MLS22 EOSes. In this section, we discuss the properties of the historical SCvH95, the ab initio MH13, and the most c  urrent CMS19 EOSes and compare their properties\footnote{The MLS22 EOS result is a hydrogen EOS table benchmarked with structural calculations of the interior of Jupiter inferred from the \textit{Juno} data. Though we present this EOS and provide the same quantities and derivatives for it, we focus here on the CMS19 and SCvH95 EOSes.}. 

\subsection{Saumon, Chabrier, and van Horn (1995)}\label{subsec:saumon}

\citet[][SCvH95]{Saumon1995} employed the chemical picture and generated an EOS that covered a temperature range of $2.10 \leq \log_{10}T\ [K] \leq 7.06$ and a density range of $-6.00 \leq \log_{10}\rho\ [\rm{g\ cm}^{-3}] \leq 3.75$. They calculated pure hydrogen and helium tables and relied on the AVL, to handle a general helium mass fraction dependence. Intensive variables, such as temperature and pressure, must be uniform throughout a system at equilibrium. Extensive variables, such as volume and entropy, are additive for a system in equilibrium.  Therefore, the AVL considers a combined system of equations of state and assumes they are in thermal and pressure equilibrium. AVL cannot account for interactions between hydrogen and helium, is not reliable in regions of partial ionization, and it does not account for phase separation (see Section~6 of SCvH95 for more detailed descriptions of its limitations). Despite these limitations, until recently, SCvH95 was the most used EOS in the study of low-mass stars, brown dwarfs, and giant planets \citep[e.g.,][]{Burrows1995, Guillot1996, Saumon1996APlanets, Burrows1997, Chabrier1997, Baraffe1998, Burrows2001, FortneyHubbard2003, FortneyHubbard2004, Fortney2011}

The additive volume law states that
\begin{equation}\label{dens_avl}
    \frac{1}{\rho (P,\ T)} = \frac{1 - Y}{\rho_{\rm{H}} (P,\ T)} + \frac{Y}{\rho_{\rm{He} } (P,\ T)},
\end{equation}
where $\rho_{\rm{H}}$, $\rho_{\rm{He} }$, and $\rho$ are the mass densities of hydrogen, helium and mixtures, respectively, at constant pressure and temperature. This corresponds to
\begin{equation}\label{ent_avl}
    S(P,T) = S_{\rm{H}}(P, T) (1 - Y) + S_{\rm{He}}(P, T)Y + S_{\rm{id}},
\end{equation}
where $S_{\rm{H}}$, $S_{\rm{He}}$, and $S_{\rm{id}}$ are the hydrogen, helium, and the ideal entropy of mixing, respectively. The SCvH95 EOS estimated entropy of mixing for each component $i$ of $m$ systems,
\begin{equation}
    S_{\rm{id}} = k_b \bigg[N\ln{N} - \sum_{i=1}^m N_i\ln{N_i}\bigg].
\end{equation}

This generalizes to all ionization states in their EOS and was used to compute the ideal entropy of mixing for H-He mixtures (see equations 53--56 in SCvH95).

The SCvH95 EOS conveniently provides many of these derivatives, making it the standard for H-He EOSes in the warm-dense region of planetary interiors and low mass stars \citep{Burrows1997, Burrows2001}. For all its strengths, the most challenging region for the SCvH95 EOS is that of partial ionization in the region of $\sim{1}$ Mbar, which is relevant for gas giant interiors. The limits of the chemical paradigm, such as the ionization states of different chemical species, are overcome by ab initio simulations under the physical picture.

\subsection{Militzer \& Hubbard (2013)}

The \citet[][MH13]{Militzer2013} EOS spans a temperature range of $500 \leq T \leq 120,000$ K and a density range of $0.1 \leq \rho \leq 6$ g cm$^{-3}$. As their model is a physical model, they avoid the limitations of the low-temperature and high-pressure regime inherent in the chemical model of SCvH95. They use Density Functional Theory Molecular Dynamics (DFT-MD), coupled with the thermodynamic integration (TDI) technique, to obtain the entropy directly through the difference in Helmholtz free energy between hydrogen and helium mixtures while using the Perdew-Burke-Ernzerhof \citep[][PBE]{Perdew1996} exchange-correlation functional. To account for interactions between species and electrons via TDI, they introduced classical intramolecular pair potentials to match their DFT calculations as closely as possible \citep{Militzer2013}. 

Since MH13 calculated their ab initio EOS only for $Y = 0.245$, they used the AVL for arbitrary mixtures. Since the AVL requires another helium fraction, the $Y=0.245$ EOS table can be combined with  an ab initio helium EOS calculated by \cite{Militzer2009}. In principle any helium EOS can be used \citep[e.g., see][who used the SCvH95 helium EOS]{Miguel2016}. The AVL can be adjusted to interpolate between any given H-He mixture, $Y_1$ and $Y_2$, by

\begin{equation}\label{dens_avl2}
    \frac{1}{\rho} = \frac{Y - Y_1}{Y_2 - Y_1}\frac{1}{\rho_2} + \frac{Y_2 - Y}{Y_2 - Y_1}\frac{1}{\rho_1},
\end{equation}  
where $Y_1 \leq Y \leq Y_2$, $\rho_1$ and $\rho_2$ are the densities at $Y_1$ and $Y_2$ respectively. For a pure helium EOS and a mixture EOS, Equation~\ref{dens_avl2} reduces to Equation 3 of \cite{HubbardMilitzer2016}. 

The MH13 EOS addressed many of the shortcomings of the SCvH95 EOS, but its lack of coverage in the shallow interior ($\lesssim 0.1$ Mbar) prompted the use of other EOS prescriptions in such regions. To overcome these obstacles, recent work \citep[see][for example]{Mankovich2021} uses a combination of EOS tables, including that of MH13, for better coverage in the density, temperature, and pressure space and greater flexibility with helium fraction dependencies. Next, we discuss the most recent work by CMS19 and CD21 to obtain a comprehensive H-He EOS.

\subsection{Chabrier, Mazevet, and Soubiran (2019)}

\citet[][CMS19]{Chabrier2019} constructed pure hydrogen and helium EOS tables spanning wider density and temperature ranges, namely $10^{-8} \leq \rho \leq 10^6$ g cm$^{-3}$ and $100 \leq T \leq 10^8$ K. For densities lower than 0.05 g cm$^{-3}$, they use the SCvH95 hydrogen EOS, the hydrogen EOS of \cite{Caillabet2011} for $0.3 \leq \rho \leq 5.0$ g cm$^{-3}$, and the EOS of \cite{Chabrier1998} for densities greater than 10 g cm$^{-3}$. 

As does SCvH95, their EOS relies on the AVL for helium mixtures and includes only the ideal entropy of mixing. CD21 incorporate the non-ideal interaction effects of MH13 into the CMS19 EOS in a manner similar to \cite{Miguel2016}. Whereas \cite{Miguel2016} did not account for the non-ideal entropy contributions from MH13, CD21 did. Using the AVL, CD21 obtained an ``effective'' hydrogen EOS from MH13 using the CMS19 helium EOS to obtain a mixture EOS that agrees with MH13 at $Y=0.245$ (see Figure 9 of CD21).

To further address this shortcoming, \citet[][HG23]{Howard2023} extended the interaction correction to arbitrary $Y$ by prescribing a non-ideal correction term of form

\begin{equation}\label{avl_ent_mod}
    S = S_{\rm{H}} (1 - Y) + S_{\rm{He}}Y + Y(1 - Y)\Delta S_{XY},
\end{equation}

\begin{equation}\label{avl_dens_mod}
    \frac{1}{\rho} = \frac{1 - Y}{\rho_{\rm{H}}} + \frac{Y}{\rho_{\rm{He}}} + Y(1 - Y)\Delta V_{XY},
\end{equation}
and requiring that these corrections agree with MH13 at $Y=0.245$ and with CD21 at $Y=0.275$. The corrections, $\Delta S_{XY}$ and $\Delta V_{XY}$, are assumed to be proportional to the products of hydrogen and helium mass fractions.

\begin{figure}
\centering
\includegraphics[width=0.47\textwidth]{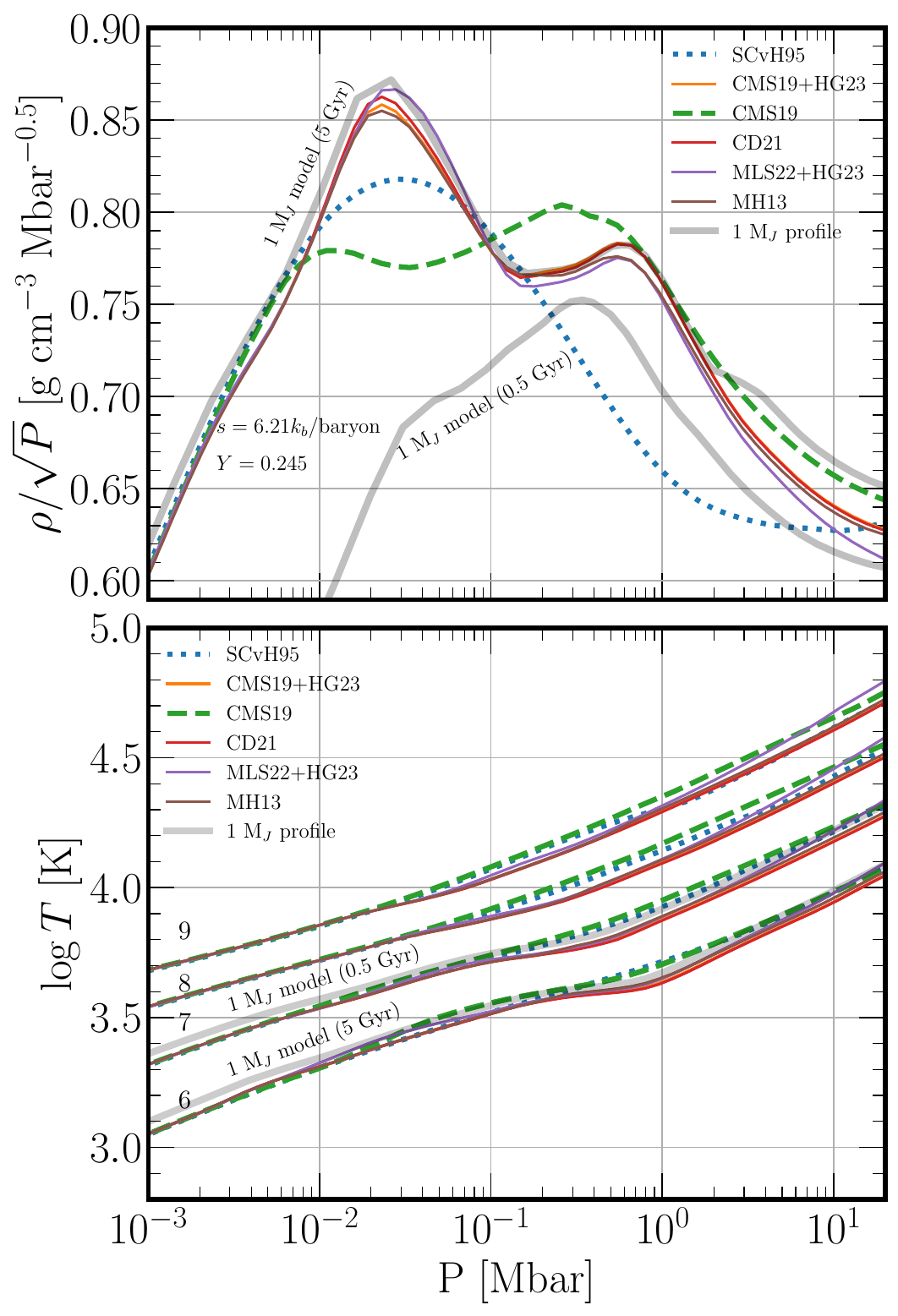}
\caption{Comparison of the H-He EOSes discussed in this study.  Top: Variant of Figure 2 of \cite{Sur2024}, originally inspired by Figure 4 of \citet{Howard2023} with corresponding labels for $Y=0.245$ and $S = 6.21\, {k_b}$baryon$^{-1}$ and with a comparison to profiles at two epochs of an evolutionary model. Solid lines indicate EOSes calculated with non-ideal interaction terms. The dotted blue line highlights the classical SCvH95 EOS, and the green dashed line the original EOS of CMS19, which relies on idealized mixing. Here we include an adiabatic 1 M$_J$ model with only H-He in the envelope, containing 10 M$_\oplus$ in a compact core. This model was calculated using \apple \citep{Sur2024} and is shown at different ages: 0.5 and 5 Gyr for reference. These profiles were calculating using the CD21 H-He EOS (red line). Bottom: Adiabatic profiles at different entropy values in $k_b\, $baryon$^{-1}$ units. We replicate the behavior shown in Figure 10 of \citet{Militzer2013}; at modest pressures of $\sim{1}$ Mbar, the non-ideal effects cause the equations of state to deviate from SCvH95. The region where the SCvH95 deviates significantly is the same region where hydrogen and helium are expected to demix. The H-He 1 M$_J$ profiles shown in the top panel are also marked in the bottom panel for reference.}
\label{fig:fig1}
\end{figure}

The CMS19 EOS with the HG23 corrections (CMS19+HG23) and the CD21 EOS are the most current development of an H-He EOS that spans all relevant thermodynamic space, helium fractions, and nonideal interactions. In Figure~\ref{fig:fig1}, we compare the CMS19+HG13, CD21, and MLS22+HG23 EOSes with the SCvH95, MH13, and CMS19 tables, illustrating the differences among them.  The top panel mirrors ~2 of \cite{Sur2024}, originally inspired by HG23, and compares the EOSes at a helium fraction of $Y=0.245$ and a specific entropy of $S = 6.21\ k_b/$baryon. The bottom panel shows the temperature differences due to the non-ideal interactions at different entropy values, as noted above, for each profile. The density deviations from SCvH95 increase with decreasing entropy, particularly between 0.5 and 7 Mbar. This pressure region is expected to exhibit H-He phase separation at a sufficiently low temperature \citep{Stevenson1975, StevensonSalpeter1977a}. Figure~\ref{fig:fig1} also compares the EOS adiabats to a 1 M$_J$ profile at 0.5 and 5 Gyrs using \apple \citep{Sur2024} with the CD21 EOS. A full example of a 1 M$_J$ evolution with \apple is discussed in Section~\ref{sec:model_example}.  In the next section, we discuss the current understanding of H-He immiscibility in this context.

\section{Hydrogen and Helium Immiscibility}\label{sec:H_He_misc}

Early evolutionary calculations for Saturn \citep{Pollack1976, Pollack1977} could not reproduce its current heat flux. To account for this apparent excess heat, \cite{Stevenson1975} proposed the formation of He-rich droplets that rain into the interior \citep[also see][]{StevensonSalpeter1977a}. One of the first attempts to capture and apply the demixing of hydrogen and helium as a function of helium abundance was done by \cite{HDW1985}, who applied Monte Carlo simulations for a fully ionized hydrogen and helium plasma. The low-pressure regime was studied by \cite{Schouten1991} via Monte Carlo methods using pair potentials for interactions. \cite{Pfaffenzeller1995} calculated the demixing temperatures using molecular dynamics (MD) and found a demixing temperature-pressure dependence opposite that of \cite{HDW1985}. However, the demixing temperatures remained too cold for Saturn to ever reach the immiscibility regions \citep{Pfaffenzeller1995, FortneyHubbard2003}. \cite{FortneyHubbard2003} created an ad hoc phase diagram, adjusting the demixing temperatures of \cite{Pfaffenzeller1995}, to match the present effective temperature of Saturn. More recently, H-He phase diagrams have been calculated by \citet[][L0911]{Lorenzen2009, Lorenzen2011}, \cite{Morales2009, Morales2013}, and \citet[][SR18]{Schottler2018}. These ab initio phase diagrams are discussed in the next sections. 

\subsection{Lorenzen et al. (2009, 2011)}

\citet[][L0911]{Lorenzen2009} conducted the first ab initio immiscibility gap model. They used DFT-MD with a PBE exchange-correlation functional and pair potentials for hydrogen and helium. They calculated the demixing temperatures for pressures between 4 Mbar and 24 Mbar and expanded their model in \cite{Lorenzen2011} to include 1 and 2 Mbar. In all cases, they relied on linear mixing to calculate the enthalpy of mixing to then infer the change in the Gibbs free energy, $G$. This was then minimized to constrain the H-He miscibility gaps. They linearized the change in the enthalpy with
\begin{equation}\label{enthalpy}
    \Delta H(x) = H(x) - xH(1) - (1-x)H(0),
\end{equation}
where $x$ is the number fraction of helium. They used the relation,
\begin{equation}\label{gibbs}
    \Delta G(x) = \Delta H(x) - T\Delta S_{\rm{id}},
\end{equation}
where $\Delta S_{\rm{id}} = -k_b[x\log{x} + (1-x)\log{(1-x)}$] is the change in the ideal entropy of mixing.

\begin{figure*}
\centering
\includegraphics[width=\textwidth]{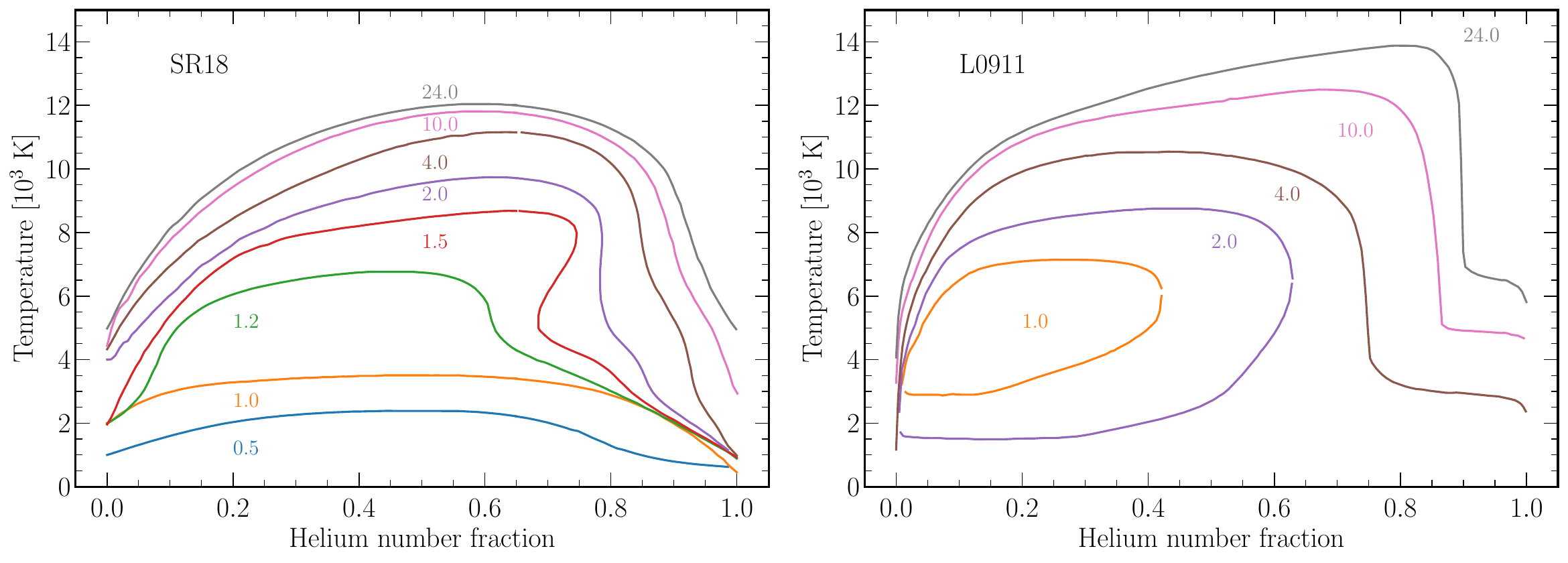}
\caption{Differences in demixing temperature calculations as a function of helium abundance. Left: \citet[][SR18]{Schottler2018}, using van der Waals exchange-correlation functional including nonideal effects on the entropy and Right: \citet[][L0911]{Lorenzen2009, Lorenzen2011}, using the PBE exchange-correlation functional and the linear mixing rule. Using the van der Waals functional and including the non-ideal effects for the entropy leads to colder demixing temperatures, particularly at low pressures. The regions under each curve are immiscibility regions of hydrogen and helium. These authors calculated such curves for constant pressures, all of which are highlighted by color here, in Mbar.}
\label{fig:fig3}
\end{figure*}

With a helium abundance-dependent phase diagram, one can interpolate between the isobars and invert it to yield the demixing temperature profiles. The right panel of Figure~\ref{fig:fig3} shows the helium dependence of the demixing temperatures for each isobar calculated by L0911. Temperatures below each of these lines indicate regions where the hydrogen and helium phase separate. In Figure~\ref{fig:fig2.5}, the SCvH95 (solid lines) and CMS19+HG23 (dotted lines) temperature profiles are calculated for a constant helium fraction of $Y = 0.27$ and a specific entropy of $S = 6.21\ k_b/$baryon, believed to be the present-day entropy of an adiabatic Jupiter as inferred by the Galileo probe entry \citep{Seiff1997, Seiff1998}. The bottom panel shows the demixing profiles of L0911. For example, since both EOS profiles intercept the $Y = 0.27$ demixing curve (in red), it is expected that demixing occurs between 1 Mbar and 2 Mbar for SCvH and 1 Mbar and 5 Mbar for the CMS19+HG3 temperature profile.

It has been suggested by both \cite{Pustow2016} and \cite{Mankovich2016} that the unperturbed phase diagrams of L0911 are unlikely due to their high demixing temperatures. \cite{Pustow2016} found that to obtain for Saturn an atmospheric helium mass fraction between $Y = 0.06$ and $Y = 0.22$ a shift of the L0911 miscibility curve needed to range from  $+500$ K to $-1,300$ K, respectively. For Jupiter, \cite{Mankovich2016} found that to fit the observed value of $Y/(X + Y) = 0.238\pm0.005$ \citep{vonzahn1998} this median shift of $-230$ K. In the next section, we discuss the work of \cite{Morales2009, Morales2013} and their inclusion of non-ideal interactions in the phase diagram.


\subsection{Morales et al. (2009, 2013)}

\cite{Morales2009} constructed the first H-He phase diagram that included non-ideal interaction effects in the entropy of mixing, following a similar methodology to that of \cite{Militzer2009} and MH13. \cite{Morales2009} used the Born-Oppenheimer MD 
coupled with the PBE exchange-correlation functional, and used TDI to infer the entropy using the Helmholtz free energy difference (also used by MH13). \cite{Morales2013} expanded their work to pressures below 0.4 Mbar, and claimed that the linear mixing assumption leads to significant disagreements with their direct calculation methods at high helium mass fractions and high pressures.

This work led to colder demixing temperatures compared to L0911. While Jupiter had a small miscibility gap between ${\sim}1$ Mbar and ${\sim}3$ Mbar with the L0911 phase diagram, the Morales demixing temperature results were too cold to allow Jupiter any miscibility gap \citep{Morales2013}. For consistency, \cite{HubbardMilitzer2016} used the Morales phase diagram, since both the EOS and the demixing temperatures were calculated in a similar fashion. \cite{HubbardMilitzer2016} then claimed that helium rain occurs within Jupiter at an entropy value of $s \approx 6\ k_b/$baryon. However, as \cite{Mankovich2020} also discuss, the published demixing temperatures are restricted to only the solar helium number fraction of $x_{\rm{He}} = 0.08$. Given that helium is expected to redistribute inside any gas giant planet, a faithful model of the interiors of these planets that takes into account helium rain must be a function of helium abundance. In the following section, we discuss the work of \cite{Schottler2018}, which used ab initio methods, including non-ideal effects, but also as a function of helium abundance.

\subsection{Sch\"{o}ttler \& Redmer (2018)}\label{subsec:SR18}

\begin{figure}
\centering
\includegraphics[width=0.45\textwidth]{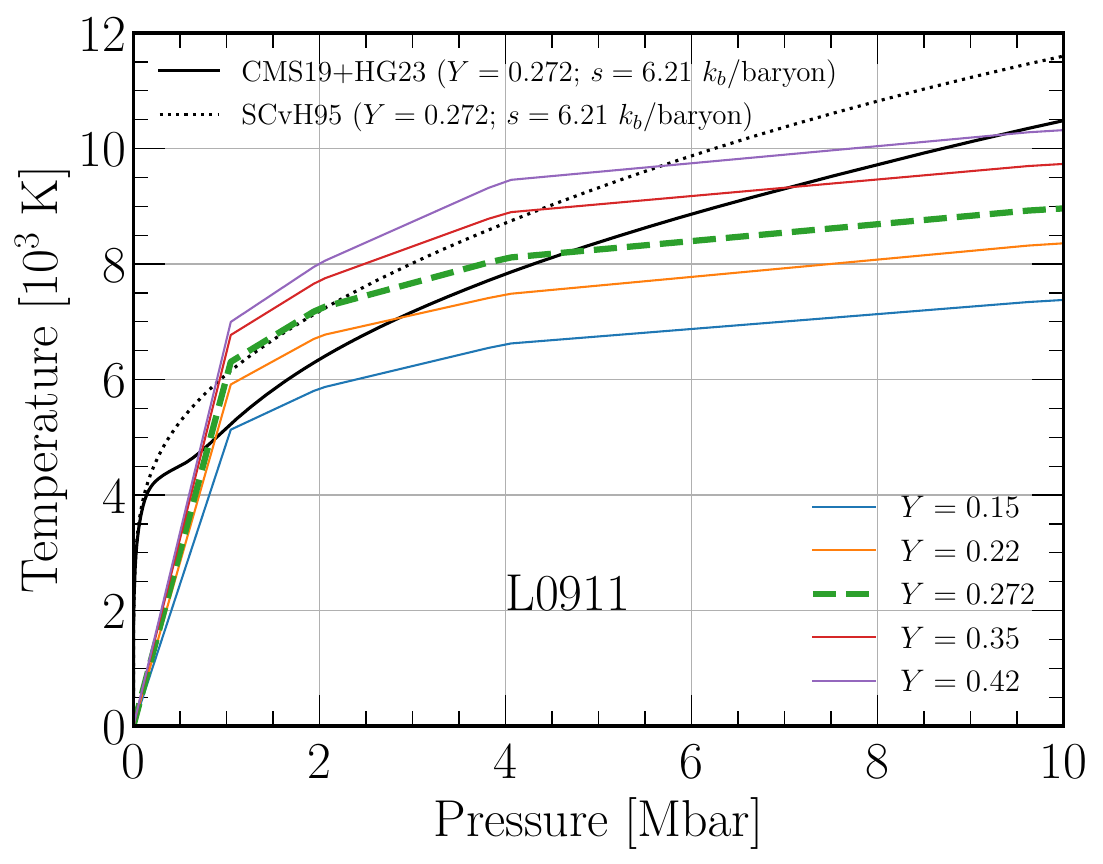}
\includegraphics[width=0.45\textwidth]{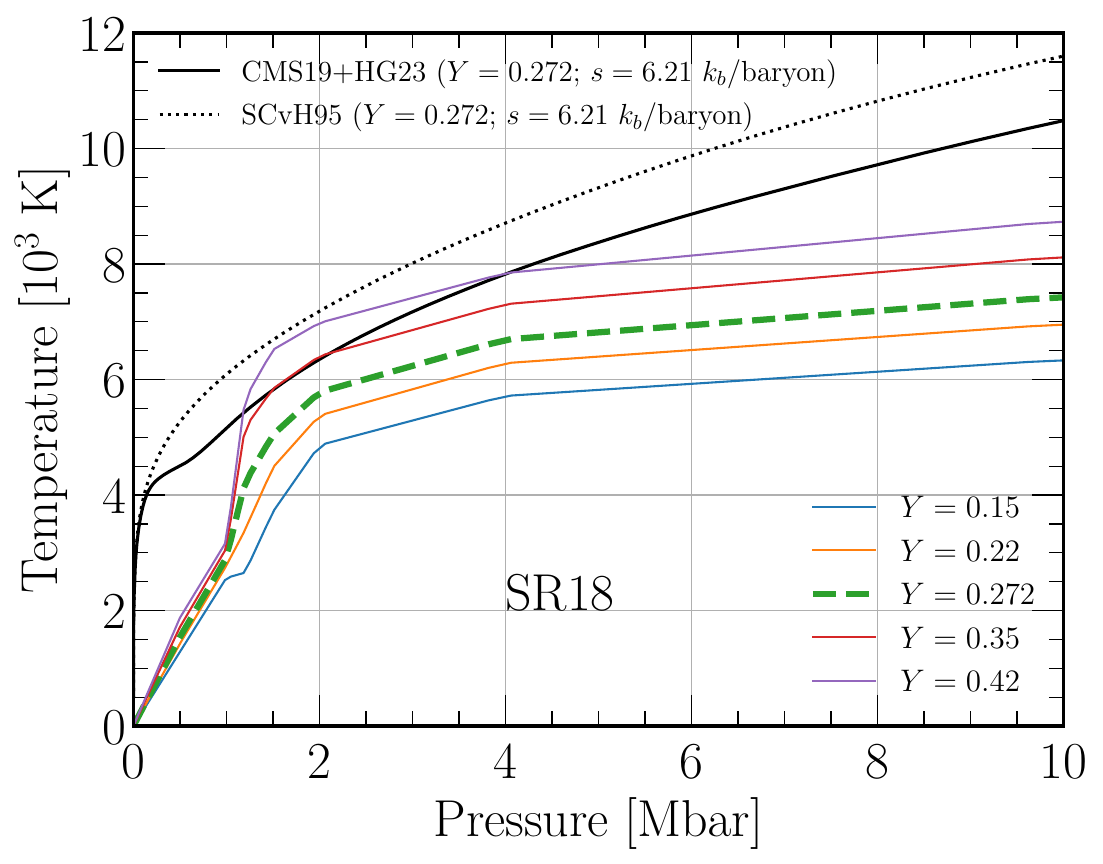}
\caption{H-He miscibility curves for various Helium fractions with SCvH95 and CMS19+HG23 adiabats at $S = 6.1\ k_b/$baryon and at a constant $Y = 0.272$. The profiles would experience H-He demixing when they intercept the $Y=0.272$ curve, which is highlighted as the dashed green line. The top panel shows curves of constant helium abundance for the L0911 data and the bottom for the SR18 data. The SR18 $Y=0.272$ curve is too cold for any of the adiabats to cross it. The L0911 $Y=0.272$ is hot enough to allow each adiabat a miscibility region. We note both that L0911 calculated the demixing temperatures down to 1 Mbar and SR18 calculated their demixing temperatures down to 0.5 Mbar. As such, the demixing temperatures below these limits shown here are extrapolations for demonstration purposes. Low pressure ($< 1.0$ Mbar for L0911 and $< 0.5$ Mbar for SR18) demixing temperature extrapolations should not be applied to evolutionary calculations.}
\label{fig:fig2.5}
\end{figure}

\citet[][SR18]{Schottler2018} performed ab initio simulations using a van der Waals exchange-correlation functional instead of PBE, claiming that their method better captured the Gibbs free energy compared to previously used methods. By calculating the non-ideal effects in the entropy, SR18 calculated the demixing temperatures shown on the left panel of Figure~\ref{fig:fig3}. It is clear that each isobar from SR18 is significantly colder than those of L0911. The 1 Mbar demixing region is significantly smaller for the van der Waals DFT result. Qualitatively, whereas L0911 showed enclosed regions of miscibility at low pressures, SR18 show flat and open regions at those pressures, leading to even colder demixing temperatures. The top panel of Figure~\ref{fig:fig2.5} shows the demixing temperature profiles with colors corresponding to those of L0911 in the bottom panel. Note that for very low helium number fractions, the temperatures are comparable, but remain significantly lower for higher helium abundances. 

\begin{figure}[ht!]
\centering
\includegraphics[width=0.45\textwidth]{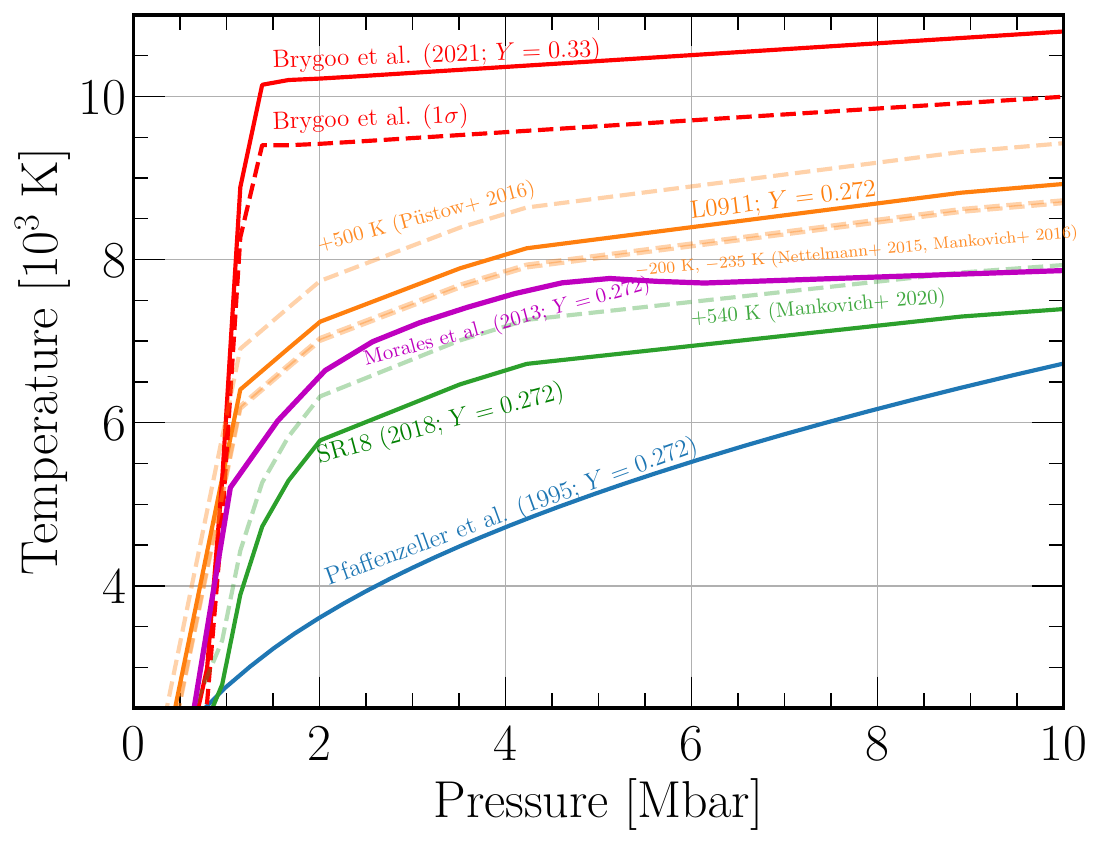}
\includegraphics[width=0.45\textwidth]{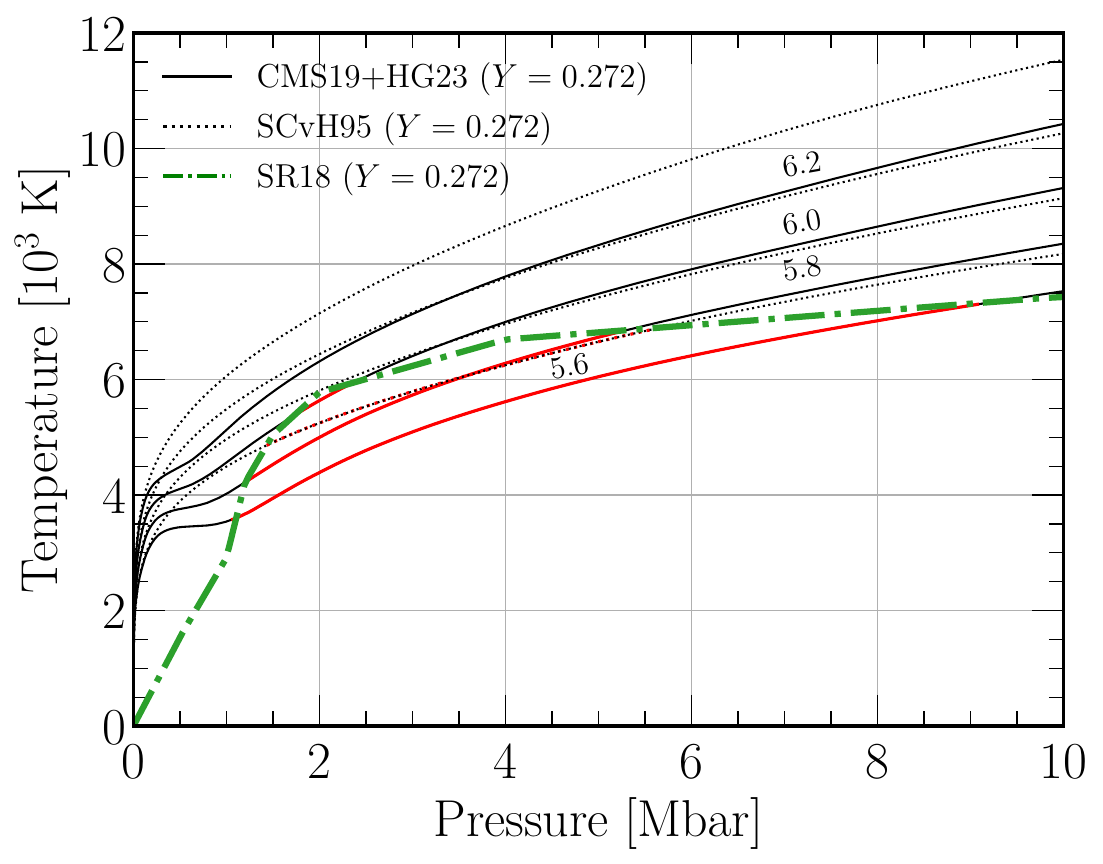}
\caption{Top: Miscibility curves from \cite{Lorenzen2009, Lorenzen2011, Morales2013, Schottler2018}, with the parametrization of \cite{Pfaffenzeller1995} from \cite{FortneyHubbard2003} at constant helium abundance ($Y = 0.272$). The dashed lines denote the temperature shifts predicted by various authors as indicated on the plot: \cite{Nettelmann2015, Pustow2016, Mankovich2016, Mankovich2020}.  The demixing temperatures of all the simulations differ from the experimental results of \cite{Brygoo2021} by more than $1\sigma$. Included is also the (minus) 1$\sigma$ error line for the \cite{Brygoo2021} data in dashed red. The experiments of \cite{Brygoo2021} were performed at $Y=0.33$. A simpler, though similar, version of this plot is presented in Figure 3 of \citet{Sur2024}. Bottom: The SR18 curve remains at $Y = 0.272$ and EOS adiabats are plotted for a range of entropy values in units of $k_b/$baryon. Once the profiles intercept the miscibility curve, the pressure regions highlighted in red will experience H-He phase separation. As entropy decreases, the miscibility gap expands to cover a wider pressure region. The SCvH95 EOS trails the CMS19+HG23 EOS and reaches the immiscibility curve at significantly colder temperatures. 
}
\label{fig:fig4}
\end{figure}

\begin{figure*}[ht!]
     \centering
    \includegraphics[width=0.45\textwidth]{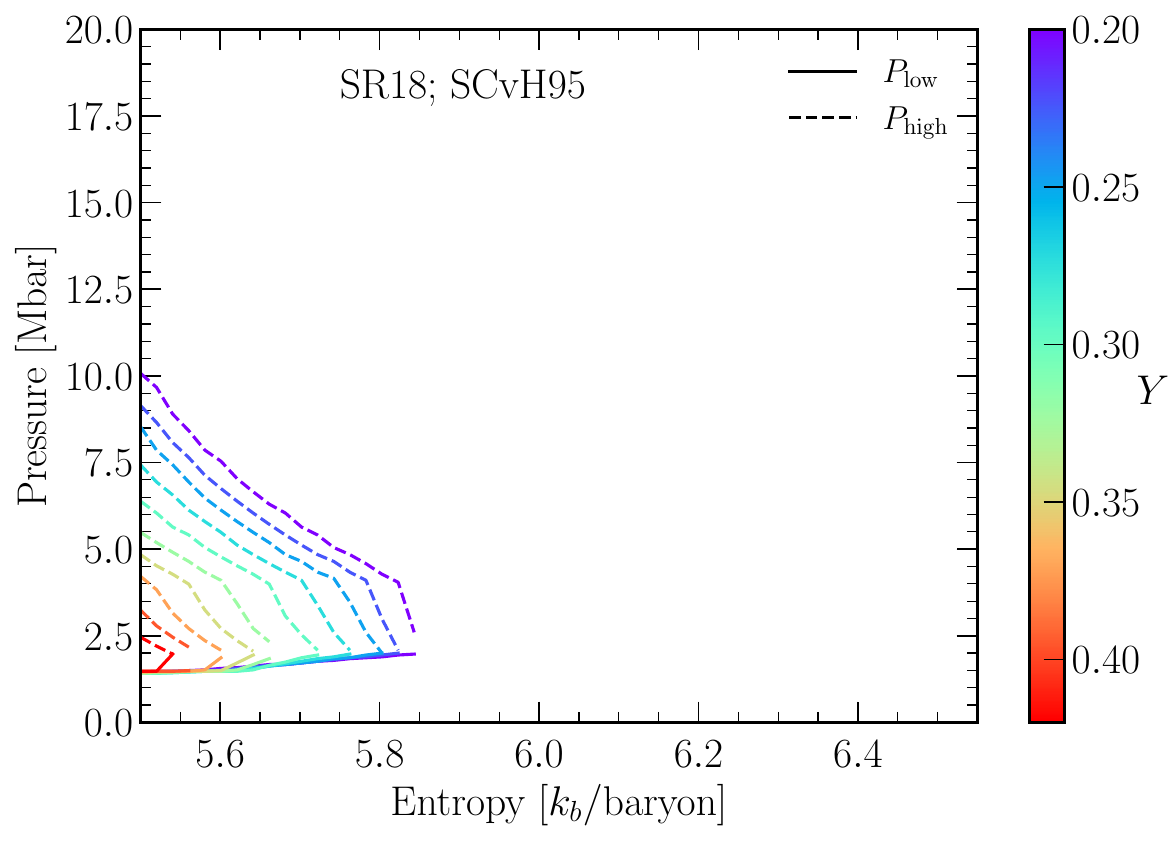}
    \includegraphics[width=0.45\textwidth]{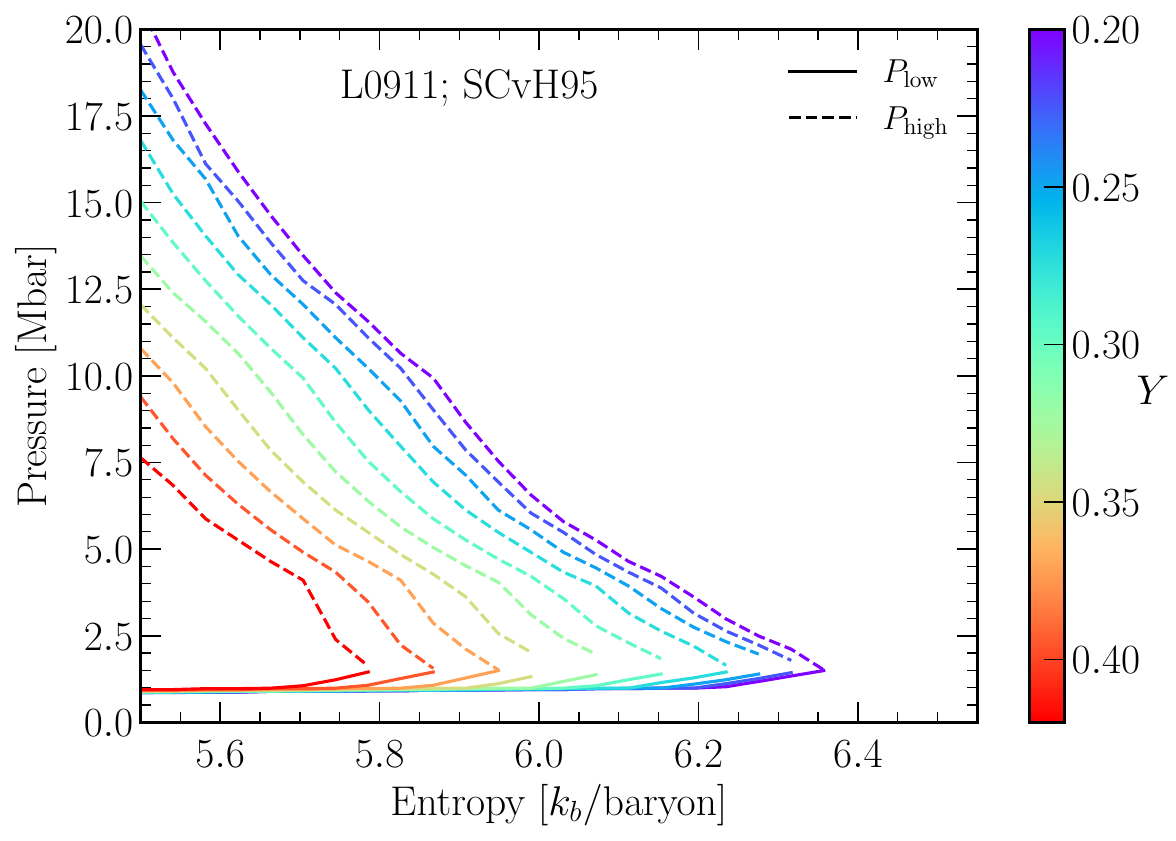}
    \includegraphics[width=0.45\textwidth]{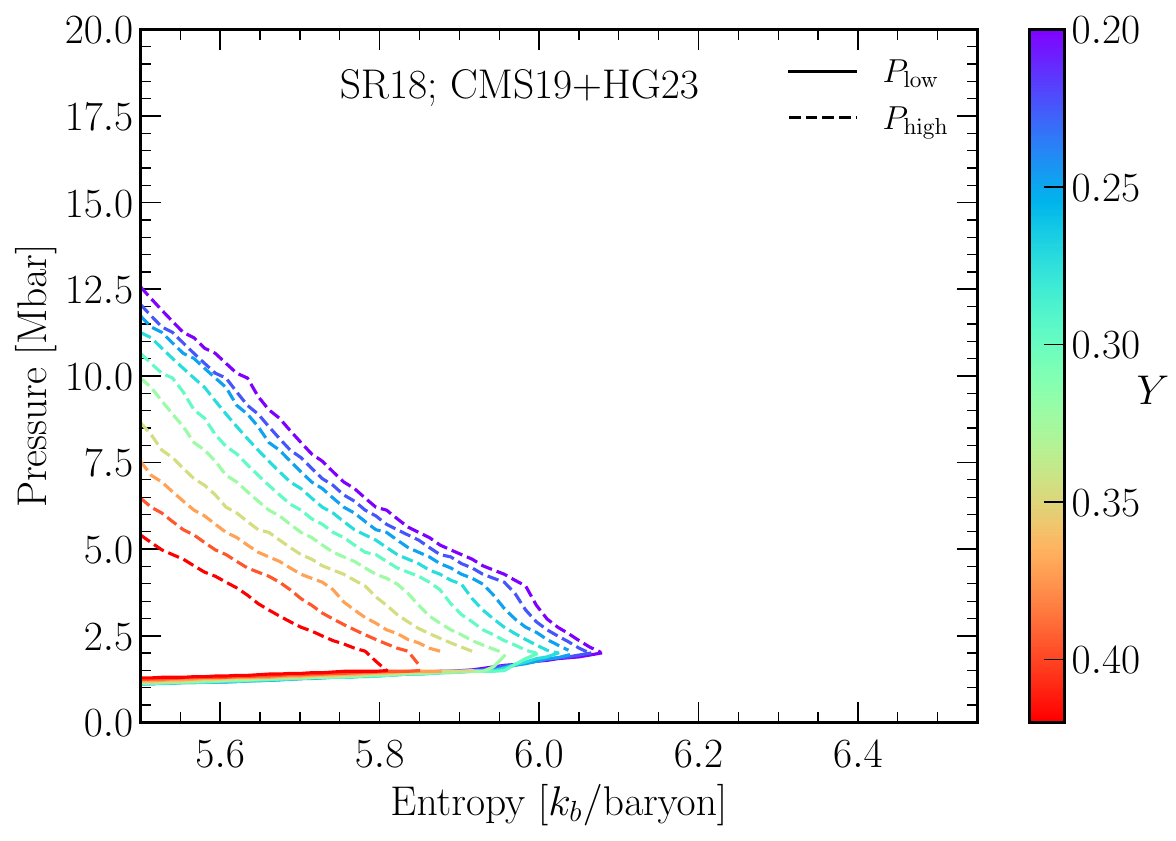}
    \includegraphics[width=0.45\textwidth]{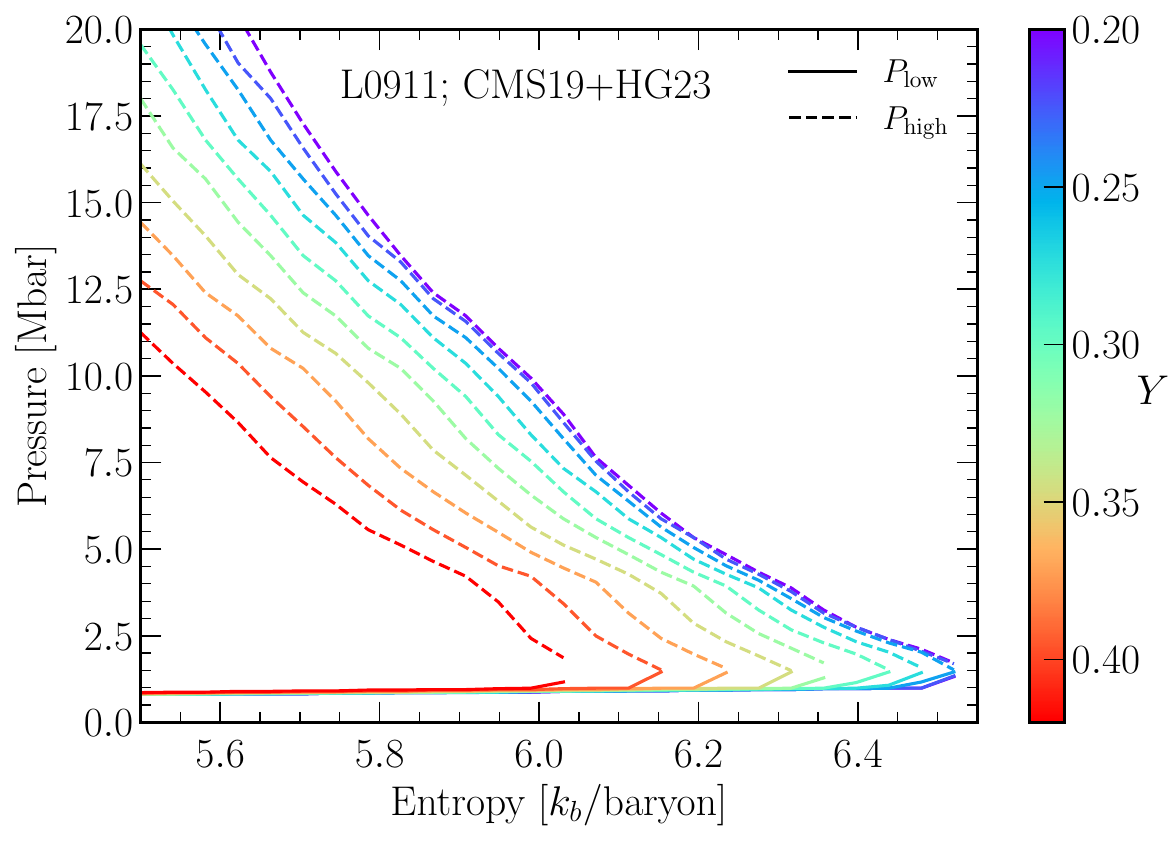}
    \caption{Pressures that enclose the miscibility gap for the L0911 (left column) and SR18 (right column) H-He immiscibility diagrams with the SCvH95 (top row) and CMS19+HG23 (bottom row) EOSes. This figure shows the low and high pressures, marked by solid lines and dashed lines, respectively, where the gap between solid and dashed lines represents the miscibility gaps as a function of entropy and helium mass fraction. These high and low pressures are the inner and outer ends of the red lines under the miscibility curve shown in the bottom panel of Figure~\ref{fig:fig4}. The low-pressure values are weak functions of helium mass fraction, a behavior also shown in Figure~\ref{fig:fig2.5}. Lower entropies are needed to initiate the immiscibility of H-He at higher helium fractions. Differences in the entropy range are due to the L0911 miscibility curves being hotter than the SR18 curves, allowing immiscibility at higher entropies.}
\label{fig:fig5}
\end{figure*}

When the demixing profiles are compared in the abundance of $Y = 0.272$\footnote{This is the primordial solar helium abundance value obtained from $Y_0/(Y_0 + X_0)$ from \cite{Bahcall2006}, where $X_0 = 0.7087$, $Y_0 = 0.2725$.}, the abundance for which Morales calculated their miscibility curves, the L0911 and Morales curves are hotter than the SR18 profiles by ${\sim} 1000$ K and ${\sim} 500$ K, respectively. The top panel of Figure~\ref{fig:fig4} shows these demixing temperature profiles (colored lines) compared to adiabats for the SCvH95 and CMS19+HG23 equations of state. If the demixing temperatures of SR18 are used to determine miscibility gaps in gas giant planets, then these will lead to narrower H-He immiscibility regions compared to L0911. The bottom panel of Figure~\ref{fig:fig4} shows decreasing entropy adiabats for both EOSes. Once the adiabat intercepts the miscibility curve, the regions highlighted in red undergo H-He phase separation. For a constant helium fraction of $Y=0.272$, this occurs at $S = 6.0\ k_b/$baryon for the CMS19+HG23 EOS and roughly at $S = 5.8\ k_b/$baryon for the SCvH EOS. We show the L0911 and SR18 miscibility gaps as a function of entropy and helium mass fraction in Figure~\ref{fig:fig5}. The helium rain regions grow as entropy and helium mass fraction decrease. As helium mass fraction increases, immiscibility regions open only at lower entropies. Figure~\ref{fig:fig5} illustrates the difference of each miscibility diagram, where L0911 can achieve miscibility gaps at higher entropies than SR18. Since the SR18 curves are significantly cooler, they require lower entropies. Because of this, \cite{Mankovich2020} found that if helium rain is responsible for the excess heat of Saturn, the H-He demixing temperatures of the SR18 miscibility curves need to be shifted by ${\sim} +540$ K. Recently, \cite{Brygoo2021} empirically measured a miscibility curve for $Y = 0.33$, highlighted in red in the top panel of Figure~\ref{fig:fig4}. While they found a positive upward shift in the temperature from the SR18 curve, the shift is ${\sim} 2,000$ K, which is far larger than the shift predicted by \cite{Mankovich2020}. 

The SR18 H-He miscibility data have several advantages over the L0911 and Morales data. Its advantage over the L0911 phase diagram is its inclusion of non-ideal entropy of mixing, while both the SR18 and L0911 data have an advantage over the Morales phase diagram in its provision of demixing temperatures for an arbitrary helium mass fraction.
We make the L0911 and SR18, and the \cite{Brygoo2021} experimental immiscibility curves conveniently accessible to use for giant planet evolutionary and structural calculations\footnote{All smooth miscibility curves are provided here: \url{https://github.com/Rob685/hhe_eos_misc/tree/main/misc}}. These H-He miscibility prescriptions are allowed to follow the pressure, temperature, and helium abundance of a given planetary interior profile. These adapt to the conditions of the interior rather than remaining static; i.e., as the pressure and temperature of the planet interior changes, so does the H-He immiscibility curve. Note that these miscibility curves were originally calculated using pure H-He mixtures. Therefore, these curves do not account for alterations of the planetary profiles due to metals. Moreover, the H-He EOSes used to calculate the immiscibility curves by SR18 and L0911 differ from those provided here.

\begin{figure*}[ht!]
\centering
\includegraphics[width=0.45\textwidth]{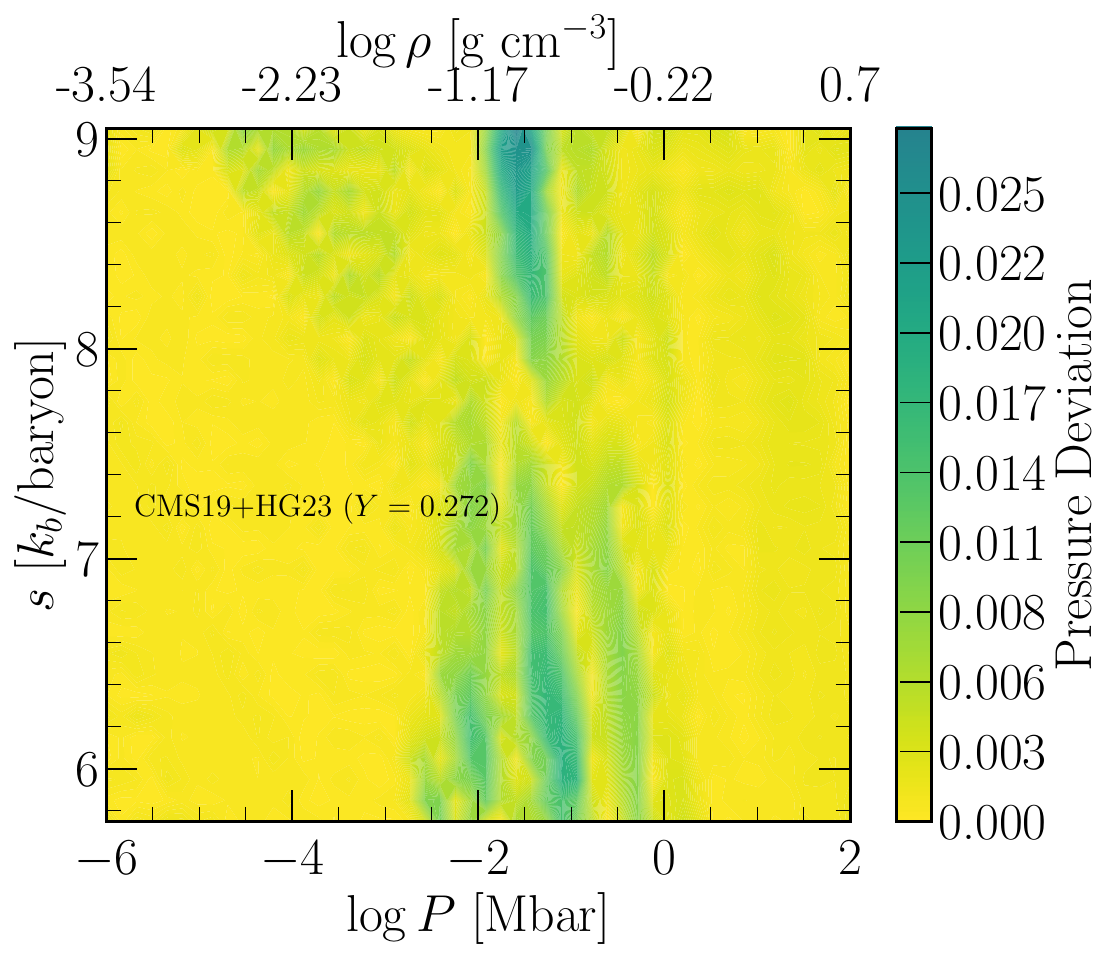}
\includegraphics[width=0.45\textwidth]{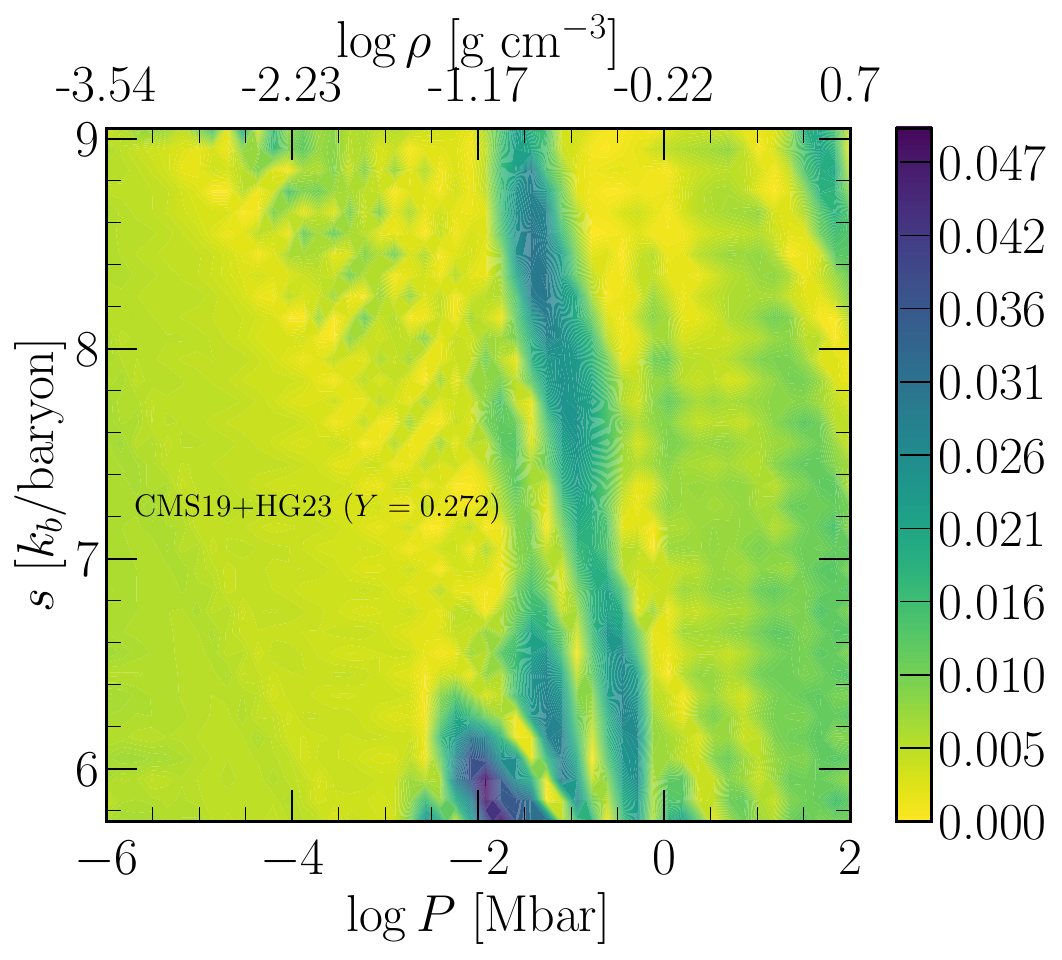}
\includegraphics[width=0.45\textwidth]{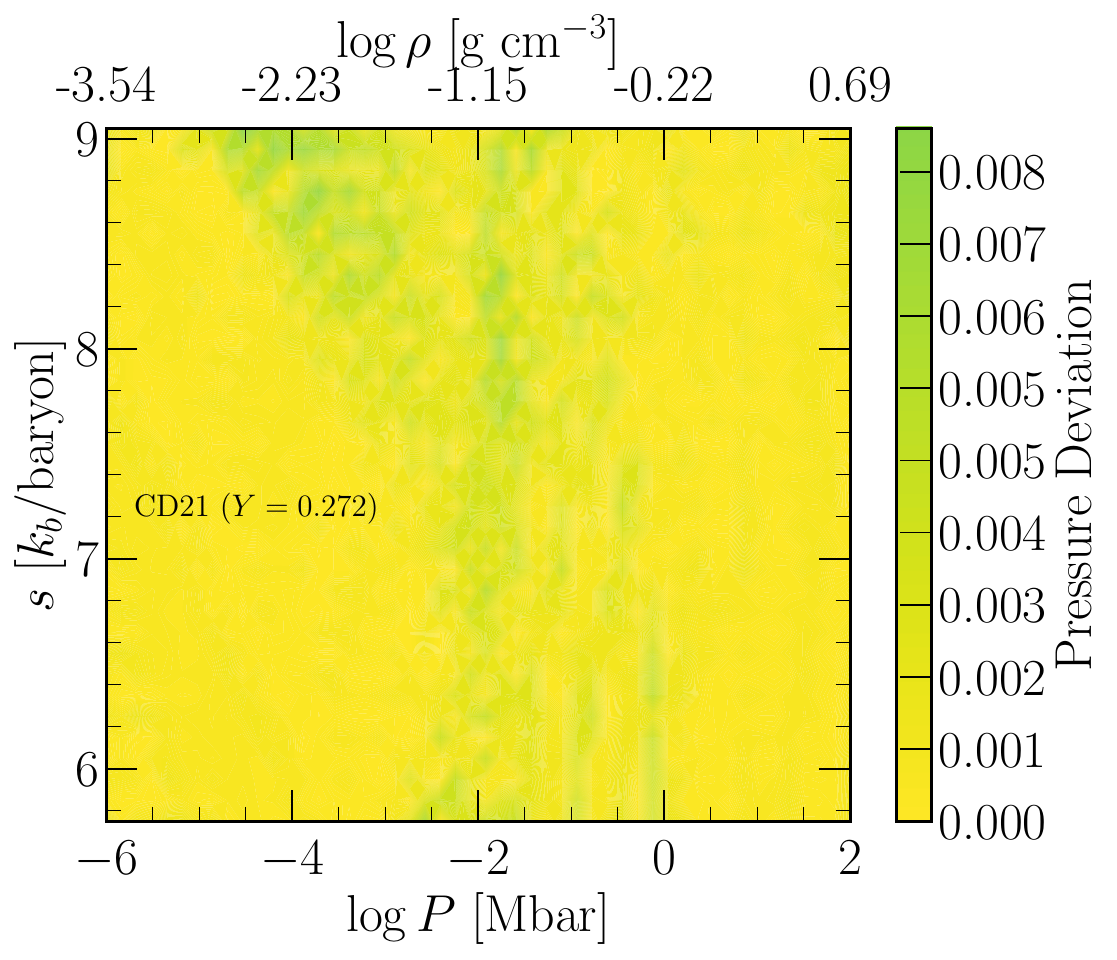}
\includegraphics[width=0.45\textwidth]{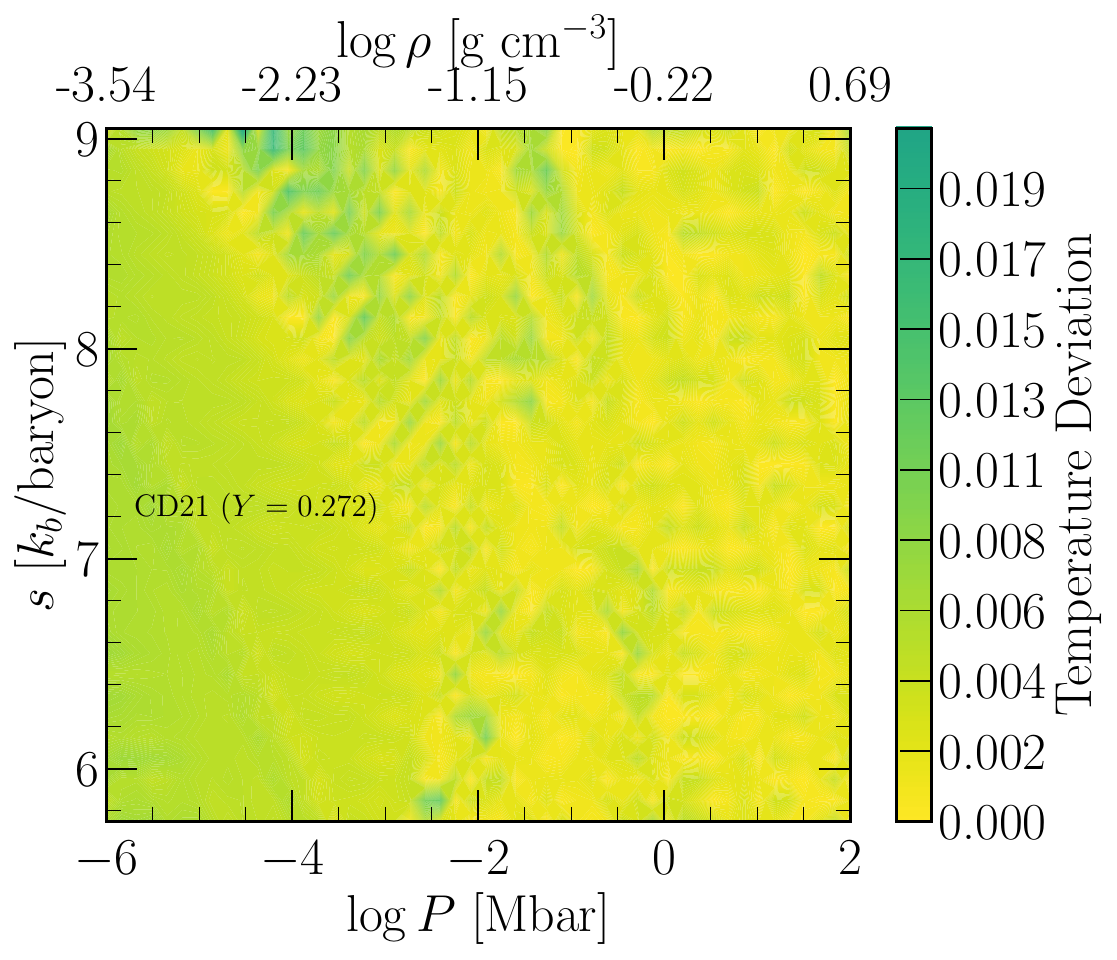}
\includegraphics[width=0.45\textwidth]{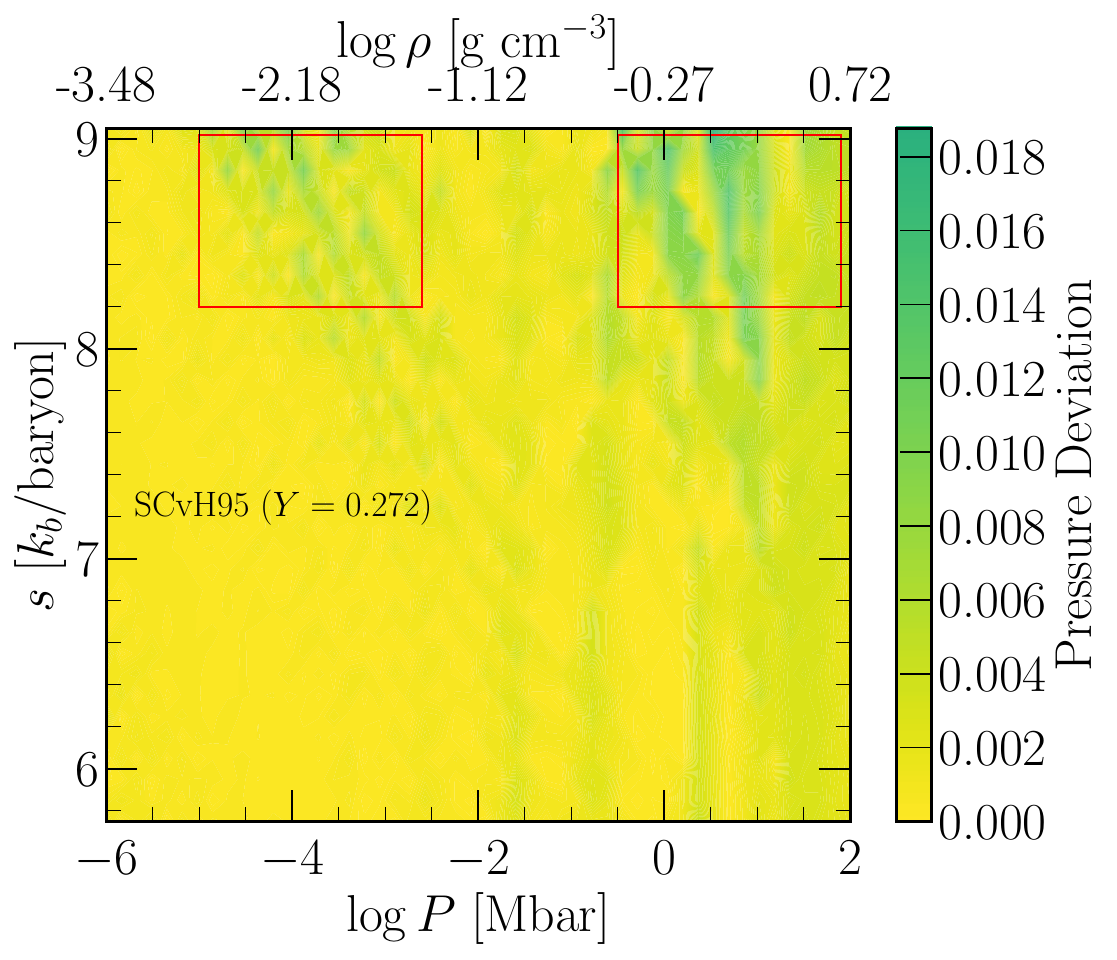}
\includegraphics[width=0.45\textwidth]{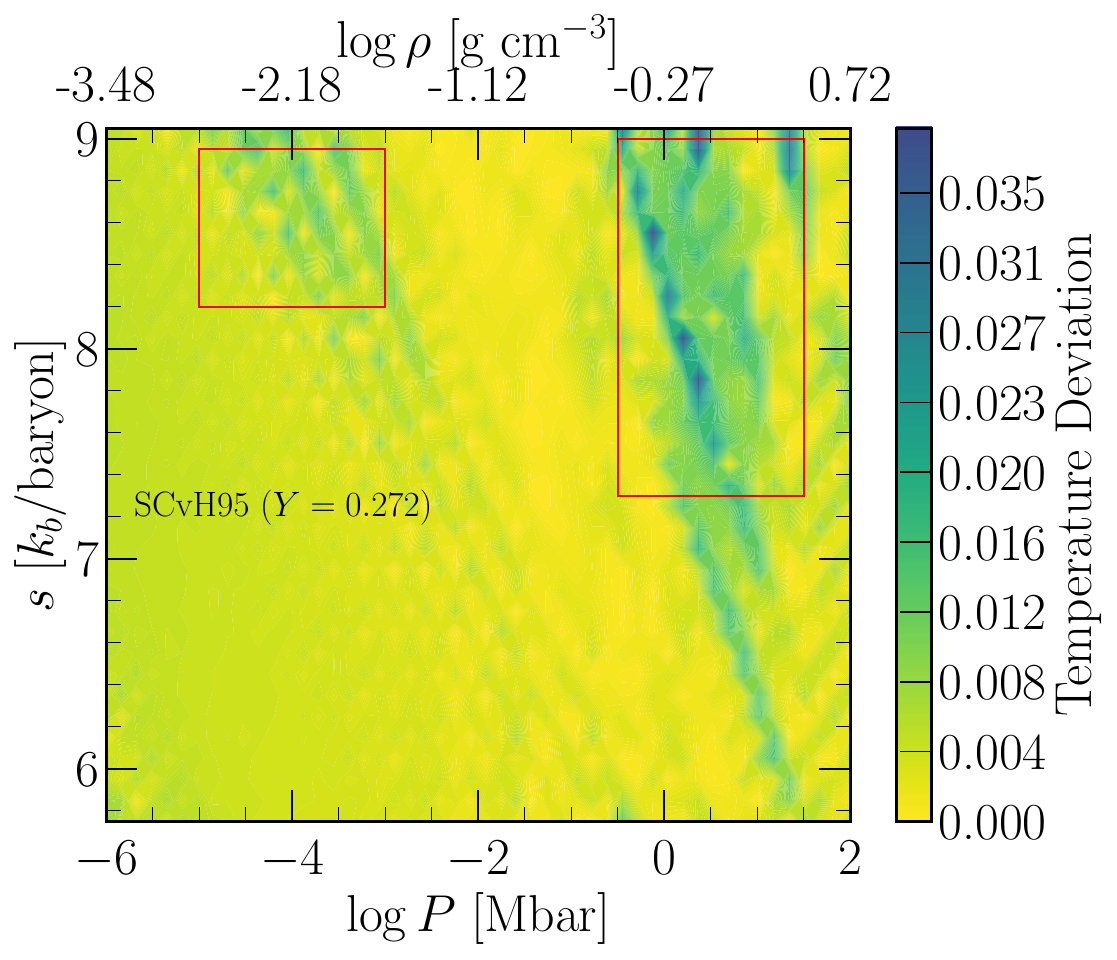}
\caption{Deviations from thermodynamic consistency indicated by Eqs. \ref{T_energy} and \ref{P_energy} for the CMS19+HG23 EOS (top row), the CD21 EOS (middle row) and the SCvH95 EOS (bottom row) at constant $Y = 0.272$. Zero indicates thermodynamic consistency. The greatest errors in the CMS19+HG23 EOS appear at low-entropy near $\sim 0.01\ \rm{Mbar}$ near the metallic hydrogen phase transition. The low entropy region corresponds to a temperature range of $\sim 2,000\ \rm{K} \leq T \leq 4,600\ \rm{K}$ at these pressures and helium mass fraction abundance. The inconsistency regions in the CMS19+HG23 EOS are largely due to the lack of non-ideal internal energy corrections, and the inconsistency regions of the SCvH95 EOS are due to the interpolation from the molecular and metallic phases to the plasma phase (see Section 3.3 of SCvH95). Such interpolated regions have been highlighted in red squares. The CD21 EOS provides internal energy values consistent with the non-ideal entropy contributions, making it more thermodynamically consistent than the rest.}
\label{fig:fig5.5}
\end{figure*}

\section{EOS \& Miscibility tables and methods}\label{sec:tables}

This work provides three additional inverted tables in addition to the original tables published by SCvH95, MH13, CMS19, CD21, and MLS22. As shown in Table~\ref{quantities}, the tables computed here have the following independent thermodynamic variable pairs: $\rho,\ T$; $S,\ P$; and $S,\ \rho$. These tables have been computed across values of $Y$ and $Z$. All table ranges are described below. While SCvH95 and CMS19 have published tables in both pressure-temperature and density-temperature, EOS mixtures of $Y$ and $Z$ using the AVL (Eqs.~\ref{avl_ent_mod} and \ref{avl_dens_mod}) can be performed only when combining entropy, density, and internal energy \citep[e.g.,][]{Saumon1995}. Tables with $\rho,\ T$ as independent variables require this density be the density of the mixture and the temperature of the mixture. Likewise, tables with $S,\ \rho$ independent variables require that the particular independent entropy and density be the mixture quantities, and not those of the original H and He tables.

To meet this requirement, numerical inversions are performed to obtain the appropriate independent thermodynamic variables; i.e., if an EOS call requires the temperature as a function of entropy, pressure, Y, and Z, then the entropy and pressure should be that of the H-He-metal mixture, and the appropriate temperature should be returned. We use a modified Powell method from MINPACK \citep{More1980UserMINKPACK-1} as applied in \textit{SciPy} \citep{Virtanen2020}\footnote{The multi-variable inversions were performed using \textit{Scipy}'s \texttt{root} application of MINPACK's \texttt{hybr} \texttt{FORTRAN} routines.} using an ideal H-He EOS as an initial guess. Such root finding methods allows us to recover original table values, and values between the original table points, to fractional errors between $\sim$$10^{-7}$ to $\sim$$10^{-15}$. All original and subsequent inverted tables are interpolated using piecewise multi-linear methods \citep{Weiser1988ADimensions, Virtanen2020}.\footnote{The 4-dimensional linear interpolations are performed via \textit{Scipy}.}

\subsection{EOS table ranges}
We list all the thermodynamic variables for each table in Table~\ref{quantities} and the units are specified in Table~\ref{units}. Each H-He EOS is combined with the pure water AQUA table \citep{Haldemann2020}. For brevity, each H-He-Z mixture table name is abbreviated with +AQUA to indicate a H-He-Z mixture. The ($S,P$) CMS19+HG23+AQUA and CD21+AQUA tables range from $4$ k$_b/$baryon to $9$ k$_b/$baryon, and the SCvH95+AQUA and MLS22+HG23+AQUA ($S,P$) tables range from $3$ k$_b/$baryon to $9$ k$_b/$baryon. All tables range from $Y = 0.05$ to $Y = 0.95$ and from $Z = 0$ to $Z = 1.0$. The ($\rho, T$) tables cover the original temperature ranges, except for temperatures above $10^5$ K.
\footnote{All the EOS tables can be extrapolated linearly from the bounds indicated, but extrapolation is not advised. The H-He EOS tables are square in shape and are designed to cover the thermodynamic space of gas-giant planet evolution, but are not yet designed to cover temperatures above $10^5$ Kelvin and densities above 100 g cm$^{-3}$.}

\subsection{H-He and Miscibility Package description}

Each miscibility curve has its own module:

\begin{itemize}
    \item L0911: \texttt{misc\_lor.py}
    \item SR18: \texttt{misc\_sch.py}
    \item \cite{Brygoo2021}: \texttt{misc\_brygoo.py}\ .
\end{itemize}

Each of these modules contains T$_m(P, Y)$ in units of $10^3$ K and P$_m(T, Y)$ in Mbar, where T$_m$ is the demixing temperature of each curve as a function of pressure and helium mass fraction and P$_m$ is its inversion.  Each of these functions has an option to offset the demixing temperatures by an amount $\Delta T$ in Kelvin. These functions were used to compute and show the miscibility curves in Figures~\ref{fig:fig2.5} and \ref{fig:fig4}.

A wrapper module called \texttt{misc.py} calls each of these files and offers users a method to calculate intercepting points between a choice of a miscibility curve and any temperature-pressure profile to generate a miscibility region. This function is used in \texttt{APPLE}'s miscibility scheme 1 \citep{Sur2024}. The \texttt{misc.py} module also contains a way to calculate Y$_m(P, T)$, or the minimum helium abundance required for H-He immiscibility for a given pressure and temperature, as well as its derivative with respect to temperature. The Y$_m$ function is used in \texttt{APPLE}'s miscibility scheme 2.

As with the miscibility curves, and as described in the EOS repository tutorials\footnote{\url{https://github.com/Rob685/hhe_eos_misc/blob/main/eos_tutorial.ipynb}}, each H-He EOS shown in Figure~\ref{fig:fig1} has its own EOS module:

\begin{itemize}
    \item SCvH95: \texttt{scvh\_eos.py}
    \item MH13: \texttt{mh13\_eos.py}
    \item CMS19 (with or without HG23): \texttt{cms\_eos.py}
    \item CD21: \texttt{cd\_eos.py}
    \item MLS22: \texttt{mls\_eos.py}
\end{itemize}
as well as the Z tables:

\begin{itemize}
    \item Water (AQUA): \texttt{aqua\_eos.py}
    \item Post-perovskite: \texttt{ppv\_eos.py}
    \item Iron: \texttt{fe\_eos.py}
    \item Serpentine (ANEOS): \texttt{serpentine\_eos.py}\ .
\end{itemize}

For H-He-Z mixtures, we include a module called \texttt{mixtures\_eos.py} that combines the user's choice of H-He EOS with the user's choice of Z EOS. The post-perovskite and iron EOS tables are provided by \cite{Zhang2022} via private communication, and the serpentine EOS from the analytical equation of state \citep[ANEOS][]{Thompson1990}. All of these files contain the pre-calculated inverted tables described in this section and in Table~\ref{quantities}, the inversion functions used for each table, and the derivatives. 

\subsection{Derivatives}

All EOS modules described and listed above offer their own derivatives, but, as described in our repository derivatives tutorial\footnote{\url{https://github.com/Rob685/hhe_eos_misc/blob/main/eos_derivatives_tutorial.ipynb}}, \texttt{mixtures\_eos.py} houses all derivatives in Table~\ref{quantities} for each H-He-Z mixture. For each derivative, the respective tables shown in the column headers of Table~\ref{quantities} are called to perform either a first order or second order finite difference. The user reserves control over the finite-difference deltas. Examples of a few derivatives used in energy transport are shown in Figure~\ref{fig:fig6.5} and Figure~\ref{fig:fig7} using fractional differences of $5\%$ along constant entropies and pressures for the purpose of differentiation. The derivatives shown in Figure~\ref{fig:fig7} are used in \apple to compute the Schwarzchild criteria ($\partial S/\partial Y|_{P, T}$) and the Ledoux criteria $(\partial S/\partial Y|_{\rho, P}$). We discuss our convection convention in Section~\ref{sec:convection}. Each of the files listed in this section can be accessed by following the installation instructions in the repository page.

\section{Thermodynamic Consistency}\label{sec:consistency}

The thermodynamic consistency for any EOS requires that its temperature, pressure, and internal energy be in agreement with the first law of thermodynamics, namely
\begin{equation}\label{T_energy}
    T(s, \rho, Y) = \parderiv{U(S, \rho, Y)}{S}{\rho}{Y} 
\end{equation}
\noindent and,
\begin{equation}\label{P_energy}
    P(S, \rho, Y) = -\parderiv{U(S, \rho, Y)}{V}{S}{Y}.
\end{equation}

In Figure~\ref{fig:fig5.5}, we show regions of thermodynamic space relevant to planetary evolution with shaded regions where each EOS deviates from consistency. The top row shows the thermodynamic consistency of the CMS19+HG23 EOS, the middle row that of the CD21 EOS, while the bottom row shows the same for the SCvH95 EOS. Thermodynamic consistency is defined here as the difference between Equations~\ref{T_energy} and \ref{P_energy} and the EOS table values over a 2-D grid of entropy and density values. In the CMS19+HG23 EOS, the metallic hydrogen phase transition is less thermodynamically consistent (denoted by darker colors), with overall errors of $\lesssim 5\%$ in the temperature. In the SCvH95 EOS, less thermodynamically consistent regions are found at higher entropies, due to interpolation techniques used by SCvH95 across the molecular/metallic transition and the plasma phase transition, with overall errors of $\lesssim 4.5\%$ in the temperature. These inconsistencies are in approximate agreement with the thermodynamic (in)consistency seen by \cite{Becker2014}, and they imply that energy conservation of planetary evolution models can only be as good as their EOSes allow. Crucially, the internal energy of the CMS19+HG23 EOS does not yet account for non-ideal interactions. This inconsistency should be partially resolved once these internal energy terms become available. On the other hand, the internal energy values of the CD21 EOS already account for these non-ideal interactions, albeit at a constant helium fraction. Unlike the SCvH95 EOS, the CD21 does not interpolate across the plasma phase transition region. This accounts for the thermodynamic consistency observed in the CD21 EOS of under $2\%$. Thermodynamic consistency must be considered when accounting for energy transport and overall energy conservation in evolutionary calculations. 

\section{Convection Criteria and The Energy Equation} \label{sec:convection}

\begin{figure*}[ht!]
\centering
\includegraphics[width=\textwidth]{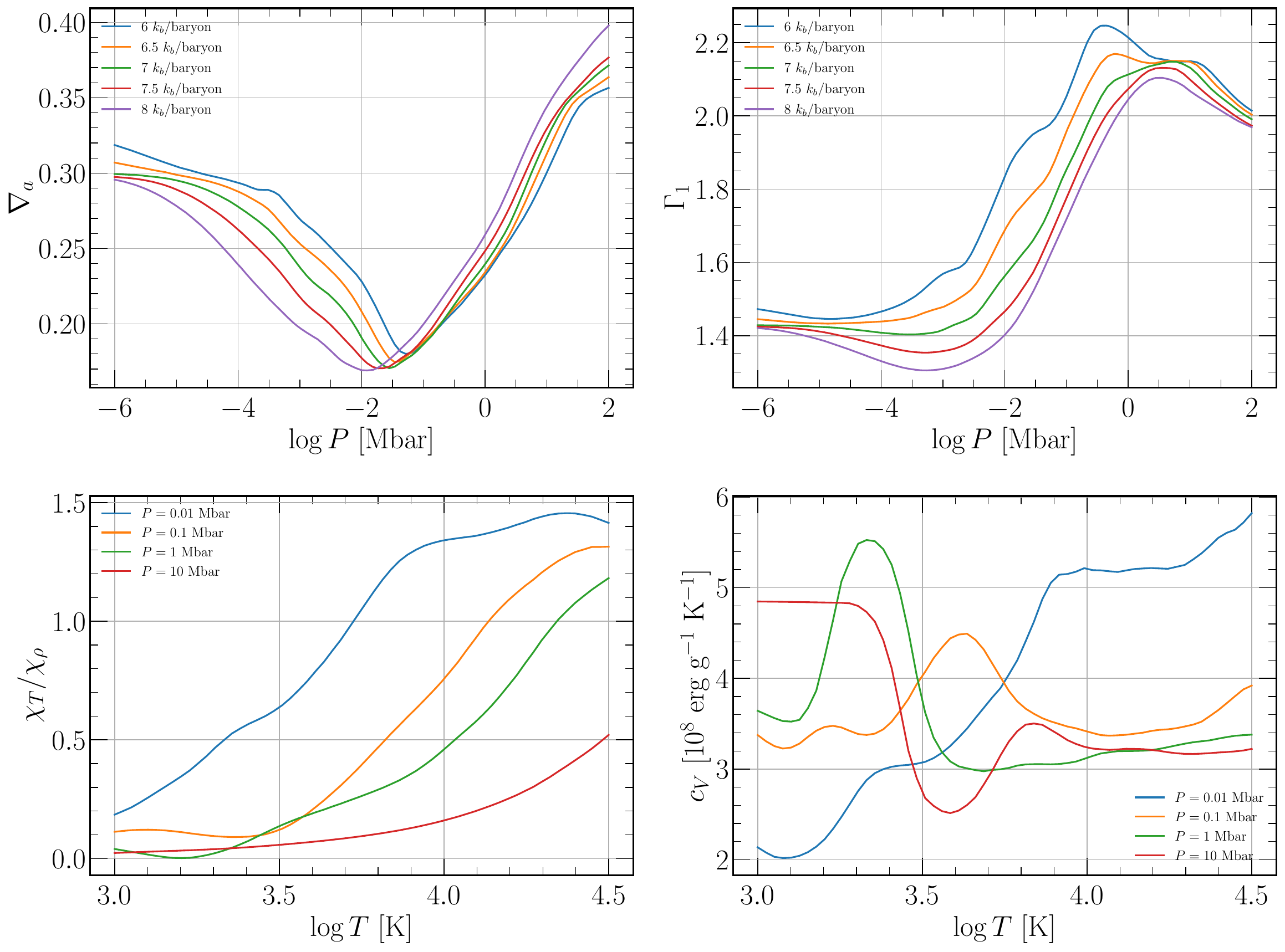}
\caption{Selection of derivative plots for the CMS19+HG23 EOS at a constant $Y=0.272$. The derivatives at constant entropy are plotted in the top row, and those at constant pressure in the bottom row. The entropy values approximately encompass standard gas giant planet evolution, and the selected pressures and temperatures for $\chi_T/\chi_{\rho}$ and the specific heat at constant volume surround the phase transition regions, highlighted in Figure~\ref{fig:fig5.5}. Note that for a non-relativistic ideal gas with no internal degrees of freedom, $\Gamma_1 = 5/3$. }
\label{fig:fig6.5}
\end{figure*}

In the standard mixing-length convection theory \citep{Cox_stellar_structure,Kippenhahn1990, Kippenhahn2012, Joyce2023}, the associated gedanken thought experiment, a fluid element is displaced and kept isolated at constant pressure with its new surroundings. Isolation in this context means the fluid element maintains a constant entropy and composition. Various structural and thermodynamic derivatives emerge to be important in the context of such theories of convection and convective transport. For simplicity, we evaluate these criteria assuming that the only compositional variation is in the helium mass fraction, $Y$, but these expressions can easily be generalized to arbitrarily complex compositional variation. The derivatives in question include: 

\begin{equation}\label{nabla}
    \nabla \equiv \logd{T}{P}
\end{equation}

\begin{equation}\label{nabla_ad}
    \nabla_{a} = \logpar{T}{P}{S}{Y}
\end{equation}

\begin{equation}\label{c_P}
    c_P = \semiparu{S}{T}{P}{Y} 
\end{equation}

\begin{equation}\label{c_V}
    c_V = \semiparu{S}{T}{\rho}{Y} 
\end{equation}

\begin{equation}
    \Gamma_1 = \logpar{P}{\rho}{S}{Y}
\end{equation}

\begin{equation}
    \gamma = \logpar{T}{\rho}{S}{Y}
\end{equation}

\begin{equation}\label{chiT}
    \chi_T =  \logpar{P}{T}{\rho}{Y}
\end{equation}

\begin{equation}\label{chirho}
    \chi_\rho = \logpar{P}{\rho}{T}{Y}
\end{equation}

\begin{equation}\label{chiY}
    \chi_Y =  \semipar{P}{Y}{\rho}{T}
\end{equation}

and 
\begin{equation}
    c_s = \sqrt{\parderiv{P}{\rho}{S}{Y}}\, .
\end{equation}
An exhaustive table of the available values and derivatives for each H-He EOSes is shown in Table~\ref{quantities}. Units used in our code are listed in Table~\ref{units}. In Figure~\ref{fig:fig6.5}, we present $\nabla_a$, $\Gamma_1$, $\chi_T/\chi_{\rho}$, and $c_V$. The entropy values selected for $\nabla_a$ and $\Gamma_1$ encompass a range relevant to gas giant planet evolution, and the pressures selected for $\chi_T/\chi_{\rho}$, and $c_V$ encompass the interior regions relevant for H-He phase separation.

\begin{deluxetable}{cccc}[ht!]
\tablecaption{List of tables available and according to each table.}
\tablehead{
\colhead{$P,\ T$} & \colhead{$\rho,\ T$} & \colhead{$S,\ P$} & \colhead{$S,\ \rho$}
}
\startdata
$\log{\rho}(P, T, Y)$ & $\log{P}(\rho, T, Y)$ & $\log{\rho}(S, P, Y)$ & $\log{P}(S, \rho, Y)$ \\
$S(P, T, Y)$ & $S(\rho, T, Y)$ & $\log{T}(S, P, Y)$ & $\log{T}(S, \rho, Y)$ \\
$U(P, T, Y)$ & $U(\rho, T, Y)$ & $U(S, P, Y)$ & $U(S, \rho, Y)$ \\
$(\partial S/\partial Y)_{P, T}$ & $\chi_T$ & $\nabla_a$ & $c_V$ \\
$(\partial \rho/\partial T)_{P, Y}$ & $\chi_{\rho}$ & $\Gamma_1$ & $\gamma$ \\
- & $\chi_Y$ & $c_P$ & $\mu_Y$ \\
- & $(\partial U/\partial \rho)_{\rho, T}$ & $c_s$ & $(\partial U/\partial S)_{S, \rho}$ \\
- & $(\partial S/\partial T)_{\rho, Y}$ & $(\partial \rho/\partial S)_{P, Y}$ & $(\partial U/\partial \rho)_{S, Y}$ \\
- & $(\partial S/\partial Y)_{\rho, T}$ & $(\partial T/\partial Y)_{S, P}$ & $(\partial T/\partial Y)_{S, \rho}$ \\
- & - & - & $(\partial T/\partial S)_{\rho, Y}$ \\
- & - & - & $(\partial T/\partial \rho)_{S, Y}$ \\
- & - & - & $(\partial S/\partial Y)_{\rho, P}$ \\
\enddata
\tablecomments{The thermodynamic quantities and derivatives are also available for pure water (AQUA), post-perovskite, serpentine, and iron, along with the provided ``metal" mixture tables\footnote{All tables are accessible at our EOS repository: \url{https://github.com/Rob685/hhe_eos_misc}}. The logarithmic quantites are log base 10, and the units are tabulated in Table~\ref{units}. The original CMS19 tables were provided by their authors in the bases $(P,T)$ and $(\rho,T)$. To ensure consistency, we used the original pressure-temperature combined with the HG23 components to obtain new $\rho, T$ tables. All quantities and derivatives of the CMS19 and MLS22 tables contain these non-ideal corrections. }
\end{deluxetable}
\label{quantities}

\begin{deluxetable}{cc}[ht!]
\tablecaption{Standard Units}
\tablehead{
\colhead{Quantity} & \colhead{Units}
}
\startdata
$S$ (independent) & k$_b/$baryon \\
$S$ (dependent) & erg g$^{-1}$ K$^{-1}$ \\
$\log{\rho}$ & g cm$^{-3}$ \\
$\log{P}$ & dyn cm$^{-2}$ \\
$\log{T}$ & K \\
$\log{U}$ & erg  g$^{-1}$
\enddata
\tablecomments{Standard units used throughout our EOS module. The EOS module provides thermodynamic quantites as a function of entropy; the independent entropy is expressed in units of k$_b/$baryon. If entropy is the dependent variable, our code will output the entropy in cgs units. We do not provide the internal energy as an independent variable.}
\end{deluxetable}
\label{units}

\subsection{Convection}
The relative density of this isolated fluid compared to that of its surroundings is the criterion for convective stability. In the absence of a composition gradient, the Schwarzschild criterion for convective stability is 

\begin{equation}\label{sch_crit_T}
    \nabla - \nabla_a < 0.
\end{equation}

In the presence of a composition gradient, the Ledoux criterion \citep{Ledoux1947} for stability must be used instead,
\begin{equation}\label{ldx_crit}
    \nabla - \nabla_a - B < 0,
\end{equation}

\noindent where $B$ is often defined as \citep{Unno1989}

\begin{equation}\label{B_T}
    \begin{split}
    B = -\frac{\chi_Y}{\chi_T} \semidu{Y}{P}. \\
    \end{split}
\end{equation}

The composition gradient term, $B$, can also be expressed in terms of entropy and composition gradients, derived in Appendix A, as

\begin{equation}\label{comp_term}
    B = -\frac{H_P}{c_P}\bigg[\parderiv{S}{Y}{\rho}{P} - \parderiv{S}{Y}{P}{T}\bigg]\fullderiv{Y}{r},
\end{equation}

where

\begin{equation}\label{B1}
    B_1 = \frac{H_P}{c_P}\parderiv{S}{Y}{\rho}{P}\fullderiv{Y}{r}
\end{equation}

\begin{equation}\label{B2}
    B_2 = -\frac{H_P}{c_P}\parderiv{S}{Y}{P}{T}\fullderiv{Y}{r}
\end{equation}
such that $B = B_1 + B_2$ and $H_P$ is the pressure scale height, $-dr/d\ln P$. The split into $B_1$ and $B_2$ is particularly useful. To whit, the superadiabaticity, $\nabla - \nabla_a$, can be shown to be
\begin{equation}
    \nabla - \nabla_a = \frac{H_P}{c_P}\fullderiv{S}{r} + B_2\, ,
\end{equation}
so the Schwarzschild criterion for convective stability can then be expressed as
\begin{equation}\label{schwarz_s}
    \frac{H_P}{c_P}\bigg[\fullderiv{S}{r}  - \parderiv{S}{Y}{P}{T}\fullderiv{Y}{r}\bigg] > 0
\end{equation}


Recall that in the absence of a composition gradient, marginal Schwarzschild stability (i.e., the transition from stability to instability) corresponds to isentropy ($ds/dr = 0$). However, Eq.~\ref{schwarz_s} shows that in the presence of a composition gradient, the transition from stability to instability occurs for a non-zero entropy gradient (even if $\nabla- \nabla_a = 0$).


Similarly,

\begin{equation}
    \nabla - \nabla_a - B = \frac{H_P}{c_P}\fullderiv{S}{r} - B_1,
\end{equation}

\noindent so that the Ledoux criterion for convective stability can be expressed as 

\begin{equation}\label{ledoux_crit_s}
    \frac{H_P}{c_P}\bigg[\fullderiv{S}{r} - \parderiv{S}{Y}{\rho}{P}\fullderiv{Y}{r}\bigg] > 0\, .
\end{equation}

In cases where a fluid is stable against convection according to the Ledoux criterion, but Schwarzschild unstable, semiconvection can arise in the form of double diffusive convection \citep[see][and references therein]{Walin1964, Kato1966, Stevenson1985, Rosenblum2011, Leconte2012, Mirouh2012, Wood2013, Leconte2013, Nettelmann2015, Moll2016, Garaud2018}. In lieu of any detailed understanding of energy transport in semiconvective regions \citep[][]{Debras2021b, Mankovich2020, Mankovich2016, Nettelmann2015, Leconte2012}, recent planetary models have included a density ratio $R_0$, which describes the degree to which temperature and composition stratifications contribute to the density stratification to trigger convection. This parameter is defined as \citep[see][for discussions]{Rosenblum2011, Mirouh2012}

\begin{equation}
    R_0 = \frac{\nabla - \nabla_a}{B}\, .
\end{equation}

Eq. \ref{ldx_crit} can be written with $R_0$ as

\begin{equation}\label{ldx_R0}
    \nabla - \nabla_a - R_0 B < 0.
\end{equation}

The parameter $R_0$ can be taken as a free parameter that stretches from 0 (Schwarzschild) to 1 (Ledoux). In terms of entropy, the criterion becomes

\begin{equation}\label{Rrho_s}
    \underbrace{\fullderiv{S}{r} - \bigg\{\parderiv{S}{Y}{P}{T}} - R_0\bigg[\underbrace{\parderiv{S}{Y}{P}{T} - \parderiv{S}{Y}{\rho}{P}\bigg]\bigg\}  \fullderiv{Y}{r}} > 0.
\end{equation}
$$\hspace{-1cm} = \frac{c_P}{H_P}(\nabla - \nabla_a) \hspace{1cm} = \frac{c_P}{H_P}(B_1 + B_2)$$
If $R_0 = 0$, then Eq. \ref{Rrho_s} reduces to the Schwarzschild criterion, Eq. \ref{schwarz_s}. If $R_0 = 1$, then Eq. \ref{Rrho_s} reduces to the Ledoux criterion, Eq. \ref{ledoux_crit_s}, modulo the $c_P/H_P$ factor.

\begin{figure*}[ht!]
\centering
\includegraphics[width=1.0\textwidth]{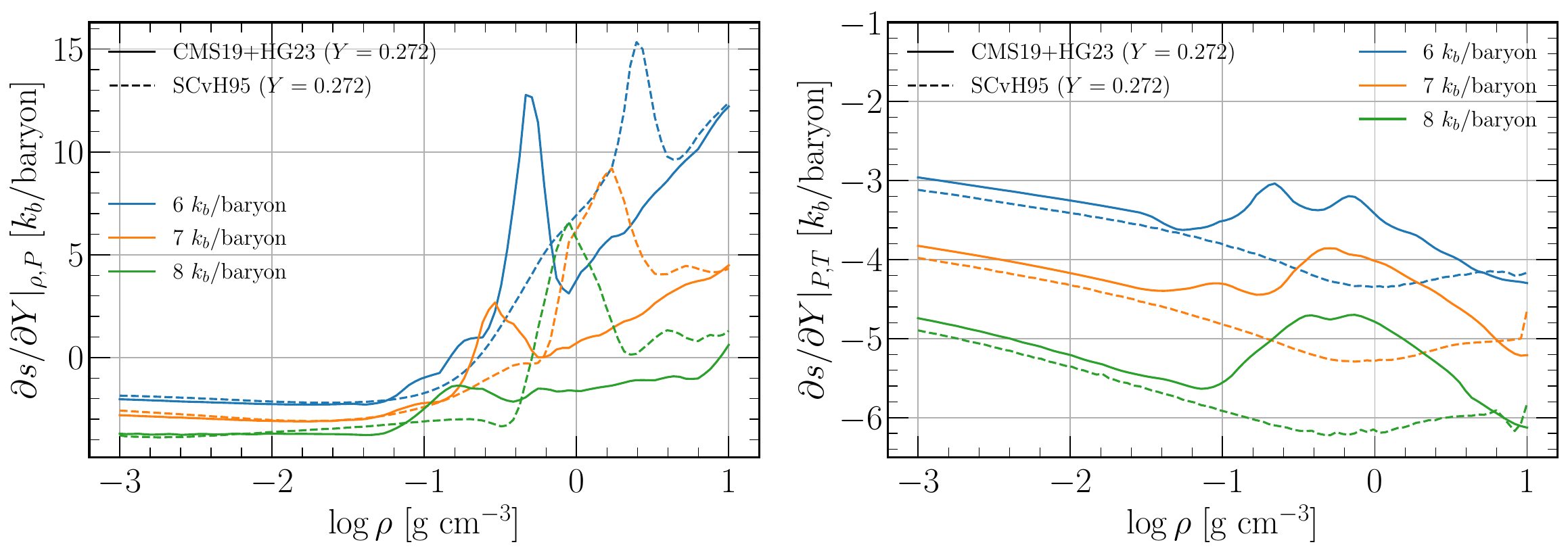}
\caption{Entropy-composition derivatives at constant $Y = 0.272$ of the CMS19+HG23 EOS (solid lines) and the SCvH95 EOS (dashed lines) as a function of density at constant $(\rho,P)$ (left) and $(P,T)$ (right). The derivative relevant to the Ledoux condition, $(\partial S/\partial Y)_{\rho, P}$, changes sign from negative to positive for densities greater than $\sim0.1$ g cm$^{-3}$. This transition occurs at temperatures $ > 1000$ K for both equations of state. The positive value of $\partial S/\partial Y |_{\rho, P}$ indicates that Ledoux-unstable regions of thermodynamic space can exhibit a negative entropy gradient, $dS/dr$ (see text).}
\label{fig:fig7}
\end{figure*}

In the repository, we provide these derivatives in functional form. Figure~\ref{fig:fig7} shows the sign changes of the relevant derivatives for the convection criteria at typical values in the interior of gas giants. Note that when the composition gradient ($dY/dr$) is negative, a negative $(\partial S/\partial Y)_{P,T}$  ensures that when the Schwarzschild convection stability condition is violated, the resulting entropy gradient is positive. This also holds for the Ledoux condition against convection, but for $(\partial S/\partial Y)_{\rho, P}$ instead of $(\partial S/\partial Y)_{P, T}$. A sign change (see Figure~\ref{fig:fig7}) of $(\partial S/\partial Y)_{\rho, P}$ at interior conditions indicates that the reverse occurs; namely, the entropy profile gradient becomes negative when the Ledoux convection stability condition is violated in these regions. The derivative at constant ($\rho, P$) increases with decreasing entropy for densities $\gtrsim 0.1$ g cm$^{-3}$.  Note also that temperatures here and for realistic gas giants are significantly higher than the hydrogen solidus line found in the CMS19 EOS. The SCvH95 EOS (dashed lines) also exhibits this behavior. This is in contrast with the Schwarzchild derivative, $(\partial S/\partial Y)_{P, T}$, which has no sign change.

We also provide the expressions to compute the Brunt-V\"{a}is\"{a}l\"{a} frequency ($N$), which is the bouyant oscillation frequency in a stable stratified fluid. This frequency is defined as \citep[][also shown in Appendix A.]{Lattimer1981, Unno1989, Saio1993, Passamonti2009, Fuller2016, Mankovich2021}:

\begin{equation}\label{brunt_rho}
    N^2 = g^2\bigg[\fullderiv{\rho}{P} - \parderiv{\rho}{P}{S}{Y} \bigg]\, ,
\end{equation}

\noindent which can be shown\footnote{in Appendix A} to be equal to 

\begin{equation}\label{brunt_S}
    N^2 = \frac{g}{c_P}\frac{\chi_T}{\chi_{\rho}}\bigg[\fullderiv{S}{r} - \parderiv{S}{Y}{\rho}{P}\fullderiv{Y}{r}\bigg].
\end{equation}

\subsection{Latent Heat Correction Due to Helium Phase Separation}
\label{latent}

 We note that the energy source that is the gravitational settling of helium should exceed by a large margin the energy source that is the latent heat of helium phase separation and that all current evolutionary codes properly handle the former. Though we emphasize that the Ledoux condition (Eqs~\ref{ldx_crit} and \ref{ledoux_crit_s}) assumes that the fluid mixtures are fully miscible, something not expected in a multi-phase context, one can approximately generalize the Ledoux condition to account for latent heat evolution (though in a miscible context) by employing what is done in the meteorological literature \citep{stevens2005} vis \`a vis wet versus dry adiabats and by employing the generalization found in
\citet{tremblin2019}.   The result is a multiplication factor on $\nabla_{\rm ad}$ (Eqs. \ref{sch_crit_T} and \ref{ldx_crit}) of the form:

\begin{equation}
\bigg[1 - \frac{\rho L}{P} \bigg(\frac{\partial{Y_{\rm m}}}{\partial{\ln P}}\bigg)_T\bigg]/\bigg[1 + \frac{L}{c_P T}\bigg(\frac{\partial Y_{\rm m}}{\partial{\ln T}}\bigg)_P\bigg] \, ,
\end{equation}
where $L$ is the latent heat per mass and Y$_{m}(P,T)$ is the helium abundance at which immiscibility ensues at a given $T$ and $P$. Assuming the latent heat estimate of \citet{StevensonSalpeter1977b} of order $\sim 0.5$\,$kT$ to $1.0$\,$kT$ per helium nucleus, the magnitude of this correction at the immiscibility curve is generally close to one, ranging from $\sim$0.93 to $\sim$0.99 for a broad range of temperature/pressure points and miscibility models. For the helium rain regions shown in Figure~\ref{fig:fig_jup_model}, it is indeed close to one. Above the immiscibility curve this correction is one and far below it one must have a better sense of the variation in the latent heat as the helium abundance associated with the phase partitioning varies. However, until complete numerical data on the latent heats under all thermodynamic and compositional circumstances are available, the helium phase-separation latent heat effect on a generalized Ledoux criterion must, for all intents and purposes, be considered unknown.

\subsection{The energy equation}
\label{sec:discussion}

The thermodynamic energy equation can be given by the first law of thermodynamics. If rendered in the independent thermodynamic basis variables ($s,\ \rho = \frac{1}{V},\ N_j$), this yields

\begin{equation}\label{erg1}
    dU + PdV = TdS + \sum^n_{j=1} \mu_j dN_j
\end{equation}
for $n$ chemical components. It is the left-hand-side of this equation that appears in the energy equation in stellar/planetary evolution and that is related to the energy sources, sinks, and the divergence of the energy flux. If the thermodynamic basis variables are ($\rho,\ T,\ N_j$), then a knowledge of $U(\rho,\ T,\ N_j)$ allows one to eschew explicit mention of $S$.  

However, if we employ the entropy formulation and include only the dependence upon $Y$ (the helium mass fraction), the specific energy (energy per mass) can be written as 

\begin{equation}\label{energy}
    dU + Pd\bigg(\frac{1}{\rho}\bigg) = TdS + \mu_Y dY\, .
\end{equation}

We note that $\mu_Y$ is not the helium chemical potential but the weighted sum emerging from the right-hand side of Eq. \ref{energy}. This can be easily expanded to include metals ($Z$) specifically by adding a $\mu_z dZ$ term, but is in fact already implicit in Eq. \ref{energy}. The composition then is included in the energy update when using entropy as an independent variable for a given evolutionary calculation step. Following Eq. \ref{energy}, the term $\mu_Y$ for the hydrogen-helium mixture is

\begin{equation}\label{mu}
    \mu_Y = \parderiv{U}{Y}{S}{\rho}\ .
\end{equation}

Recent giant planet evolutionary calculations, such as those of \cite{Mankovich2016}, and \cite{Mankovich2020}, have approximated this term by differentiating the energy throughout the resolution of the profile \citep[see Section~2.5 of ][]{Mankovich2016}. Stellar evolution codes such as MESA \citep{Paxton2011} have recently updated their fundamental energy equations to include explicitly the chemical potential \citep[see Section~8 of][]{Paxton2018}. Along with the tables provided with this article, we also provide a means of obtaining this term for any $S, \rho, Y$, regardless of the interior profile.

\section{One Jupiter Mass model examples}\label{sec:model_example}

As an example of the use of these EOS tables, we employ the evolutionary code \texttt{APPLE} \citep{Sur2024} to compute a pair of one M$_J$ evolutionary models, with and without helium rain. We assume here the CD21 equation of state and the \citet{Lorenzen2011} miscibility curves, shifted by $\Delta T = +1,000$\ K\footnote{Although this model is not designed to fit Jupiter in detail, such a shift is required to maintain a flat 3 solar metallicity envelope to closely match the observed $Y = 0.234$ abundance \citep{vonzahn1998}.}. These models do not purport to be fits to Jupiter, as a full and detailed study with \texttt{APPLE} is the subject of upcoming work (Sur et al. 2024b, \textit{in prep}). The \texttt{APPLE} code includes, as best we currently can, helium rain. In it, we employ a scale parameter ($\mathcal{H}_r$, here set equal to $10^8$ cm) that governs the magnitude of an settling/advective term, while conserving mass and properly accounting for the gravitational energy release. The parameter $H_r$ results in an equilibrium helium ``rain cloud" that is quasi-exponentially distributed in space (similar to models of condensation clouds, but consistent with the miscibility curves).

\begin{figure*}[ht!]
\centering
\includegraphics[width=1.0\textwidth]{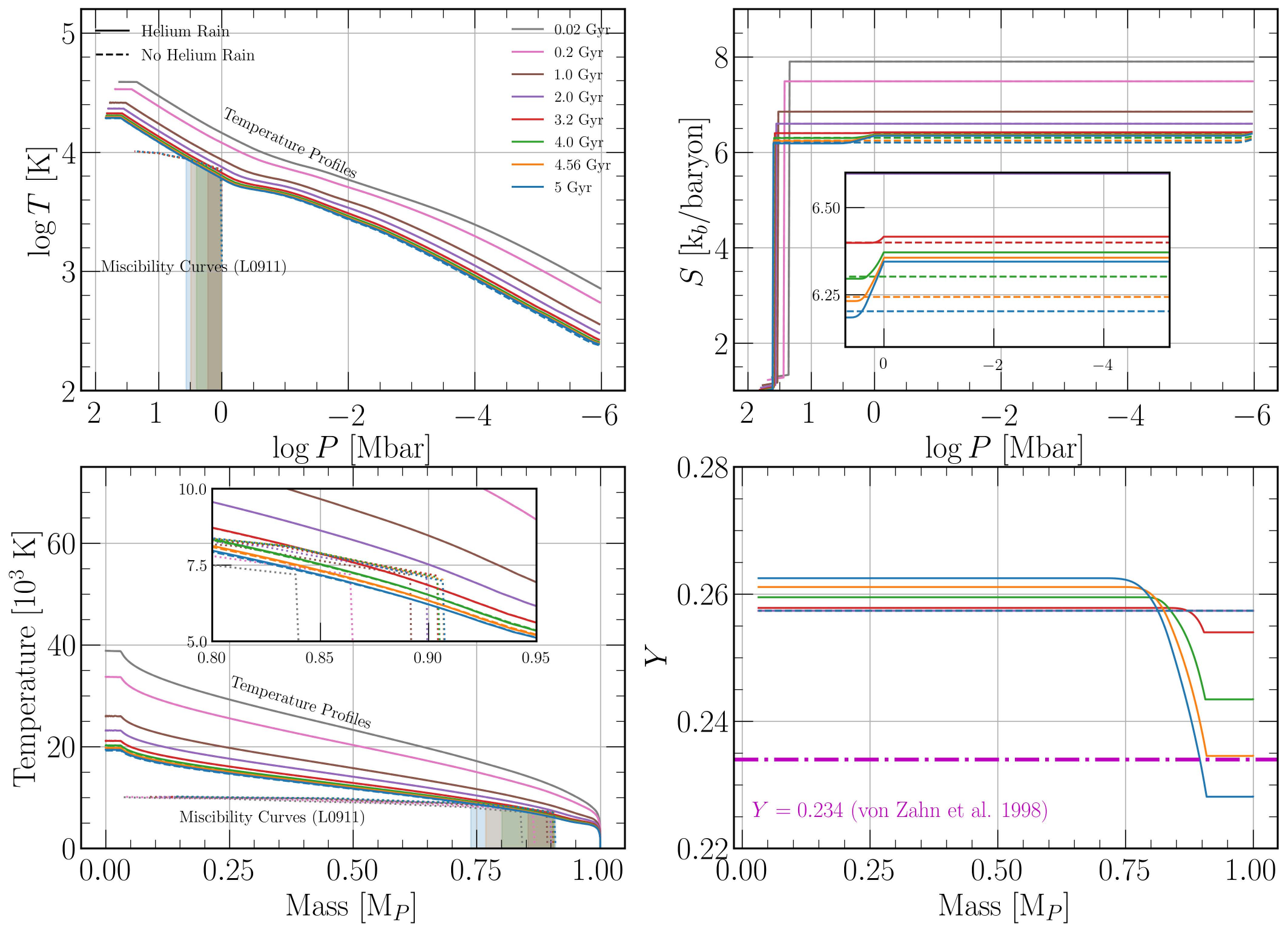}
\caption{One Jupiter mass models computed with \apple 
\citep{Sur2024}. The top left and bottom left panels show the temperature profile evolution with and without helium immiscibility (solid and dashed lines) as a function of pressure and mass. Each color represents a different time in the evolution in Gyr ranging from 0.02 Gyr to 5 Gyr. We use the L0911 miscibility curve (shown as dotted lines) shifted for this example by $\Delta T = + 1,000$ K. The adjustment to the miscibility curve to the local conditions of the profile is shown as a function of age, and the insert zooms into the region experiencing it. The miscibility curve moves closer to the profiles as the local pressure increases with age. The top right panel presents the entropy profiles. The entropy profiles drop to low values of $\sim 1$ k$_b/$baryon in the ``heavy element" core, which is $10$ M$_\oplus$ in these models. This evolution example contains 16.5 M$_\oplus$ in heavy elements. The insert focuses to distinguish the helium-rain/no-helium-rain differences. The bottom right panel shows the evolution of the total helium mass fraction profile. This particular model has been computed approximately to match the observed total helium mass fraction of Jupiter's atmosphere \citep{vonzahn1998} at its current age of $\sim$4.56 Gyrs, shown in magenta.}
\label{fig:fig_jup_model}
\end{figure*}

Figure~\ref{fig:fig_jup_model} shows the thermal profile evolution as a function of pressure and mass on the left column, and the evolution of the entropy profiles and helium profiles on the right column. These models were computed using a flat water abundance profile at 3 times solar throughout the envelope, with a total metallicity mass of $26.5$ M$_\oplus$ for both models, assuming $10$ M$_\oplus$ of ice/rock/iron mixtures in a compact core. We used atmospheric boundary conditions that included solar irradiation and ammonia clouds as computed by \cite{Chen2023}. While these models are merely test examples of the use of our tables in an evolutionary code, they do still incorporate much of the basic physics necessary to address gas giant evolution. More sophisticated models using \apple will be forthcoming.

\bigskip

\section{Conclusion}\label{sec:conclusion}

With this paper, and its associated repository, we provide a more centralized and organized way to access the crucial hydrogen-helium-metallicity mixture equations of state, accompanied by the various derivatives necessary to calculate heat transport and different convection criteria. The EOSes and derivatives are recalculated in the range $-6 \leq \log_{10}{P}\ {\rm (Mbar)}\leq 2$ and $2 \leq \log_{10}{T}\ {\rm (Kelvin)} \leq 5$, a range that is comparable to each of the EOS tables that we have adapted. Importantly, we calculate the pseudo-chemical potential (Eq. \ref{mu}) so that it is readily available for energy calculations for any helium mixture. 

Our primary purpose is to provide curated access to equation of state and miscibility data and thermodynamic derivatives presently scattered in the literature. In addition, though the standard
Schwarzchild and Ledoux conditions discussed in the stellar and planetary evolution literature \citep{Kippenhahn1990,Paxton2011,gabriel2014,mihalas1978} use temperature and composition as primary variables and refer centrally to $\nabla$ and $\nabla_{\rm ad}$, the alternate set
of variables (entropy and composition) have virtues \citep[see][]{Sur2024} and in this paper have provided the corresponding formulae.

The reader should note that the effects of radiative/conductive and composition diffusion on the Ledoux condition and the various approaches to doubly-diffusive instabilities (``semi-convection") are discussed in many works (see, e.g., \citet{Kippenhahn1990}, \citet{Paxton2011}, \citet{langer1983}, and \citet{tremblin2019}) and will not be repeated here. On doubly-diffusive instabilities, the group of Pascale Garaud has generated a series of important
papers \citep{Rosenblum2011,traxler2011,Mirouh2012,Wood2013,brown2013, Moll2016,tulekeyev} in the giant planet context.  While a full discussion
of these additional issues is beyond the scope of this paper, mention
should be made of the complicating effects of helium rain. The standard Ledoux condition doesn't explicitly incorporate the latent heat of miscibility, nor the issue of the growth and advective settling of helium droplets in a gravitational field.  Importantly, the latent heat of hydrogen/helium immiscibility for the various miscibility models is not provided in the papers that perform detailed
miscibility calculations. As a result, it is premature to delve into this topic in detail until such data from modern calculations are available. We note that the latent heat effect is likely much smaller than the heating per helium nucleus provided by the settling rain itself in a gravitational field,
and the latter effect is properly handled in extant giant planet evolution codes \citep[e.g.,][]{Sur2024}.

In addition, a study of the instantaneous effect of helium droplet settling on a generalized Ledoux condition must acknowledge that the regions of the giant planet where miscibility emerges are likely already convective and that they cool into the miscibility curve. This implies that there is already upwelling and mixing in the region for which there is incipient phase separation. This hydrodynamics is not well addressed by the standard linear stability gedanken experiment that leads to the Ledoux condition, or its standard generalizations. Moreover, mixing-length theory, the standard
context of spherical evolutionary codes, is not well suited to this multi-phase, multi-dimensional physics. Clearly, this highlights the need for a more sophisticated approach going forward to the interesting physics of helium rain in planet evolutionary models.

In summary, this work is designed to facilitate the prompt calculation of the thermodynamic quantities and their derivatives required for updated giant planet and exoplanet modeling in the \textit{Juno} and \textit{Cassini} eras and to help new groups of modelers join this exciting field.   

\begin{acknowledgments}
We thank the referee for a thorough review that very much sharpened numerous aspects of the manuscript. Funding (or partial funding) for this research was provided by the Center for Matter at Atomic Pressures (CMAP), a National Science Foundation (NSF) Physics Frontier Center, under Award PHY-2020249. Any opinions, findings, conclusions or recommendations expressed in this material are those of the author(s) and do not necessarily reflect those of the National Science Foundation. RTA is supported by the Ford Foundation Predoctoral Fellowship.
YS is supported by a Lyman Spitzer, Jr. Postdoctoral Fellowship at
Princeton University.
\end{acknowledgments}

\section*{Data Availability}
All data are available on Zenodo under an open-source 
Creative Commons Attribution license: 
\dataset[doi:10.5281/zenodo.10659249]{https://doi.org/10.5281/zenodo.10659249}.

\appendix

\section{Composition gradient term with entropy gradients}

Recall that with a composition gradient the criterion for stability against convection is Eq. \ref{ldx_crit}. Using the definition in Eq. \ref{B_T},

\begin{equation} \label{brunt_T_app}
    \nabla - \nabla_a - \semipar{T}{Y}{\rho}{P}\semidu{Y}{P} > 0\, .
\end{equation}

Recognizing that 
\begin{equation}
    \semipar{T}{Y}{\rho}{P} = \semipar{T}{Y}{S}{P} + \semipar{T}{S}{P}{Y}\parderiv{S}{Y}{\rho}{P}\, ,
\end{equation}

\noindent $B$ in Eq. \ref{B_T} can be written in terms of two derivatives of the entropy with respect to the helium mass fraction Y, 
\begin{equation}
    B = \semipar{T}{S}{P}{Y}\bigg[\parderiv{S}{Y}{\rho}{P} - \parderiv{S}{Y}{P}{T} \bigg]\semidu{Y}{P}.
\end{equation}

$$ = \frac{P}{c_P}\bigg[\parderiv{S}{Y}{\rho}{P} - \parderiv{S}{Y}{P}{T}\bigg]\fullderiv{Y}{P}$$

\noindent where the two terms can be assigned the names $B_1$ and $B_2$,

\begin{equation}
    B_1 = \frac{P}{c_P}\parderiv{S}{Y}{\rho}{P}\fullderiv{Y}{P},
\end{equation}

\begin{equation}
    B_2 = -\frac{P}{c_P}\parderiv{S}{Y}{P}{T}\fullderiv{Y}{P}
\end{equation}

\noindent such that $B = B_1 + B_2$ and as discussed in \S\ref{sec:convection}. Using the pressure scale height, these terms can be written in terms of entropy and composition profile gradients,

\begin{equation}\label{B1_prof}
    B_1= -\frac{H_P}{c_P}\parderiv{S}{Y}{\rho}{P}\fullderiv{Y}{r}
\end{equation}

\begin{equation}\label{B2_prof}
    B_2 = \frac{H_P}{c_P}\parderiv{S}{Y}{P}{T}\fullderiv{Y}{r}.
\end{equation}

\section{The Brunt V\"{a}is\"{a}l\"{a} frequency in terms of entropy and composition}

Let $\rho = \rho(S, P, Y)$ so that 

\begin{equation}
    d\log{\rho} = \logpar{\rho}{P}{S}{Y}d\log{P} + \semipar{\rho}{S}{P}{Y}ds + \semipar{\rho}{Y}{P}{S}dY\, ,
\end{equation}
\noindent implying that 

\begin{equation}
    \logd{\rho}{P} =  \logpar{\rho}{P}{S}{Y} + \semipar{\rho}{S}{P}{Y}\semidu{S}{P} + \semipar{\rho}{Y}{P}{S}\semidu{Y}{P}\, .
\end{equation}

Using the relation on the right-hand side of Eq.~\ref{brunt_rho}, 

\begin{equation}
    N^2 = \frac{\rho g^2}{P}\bigg[\semipar{\rho}{S}{P}{Y}\semidu{S}{P} + \semipar{\rho}{Y}{S}{P} \semidu{Y}{P}\bigg]\, ,
\end{equation}

\noindent which becomes

\begin{equation}
    N^2 = \frac{\rho g^2}{P}\semipar{\rho}{S}{P}{Y}\bigg[\semidu{S}{P} - \semiparu{S}{Y}{\rho}{P}\semidu{Y}{P} \bigg]\, .
\end{equation}

Using 

$$g^2 = -\frac{gP}{\rho}\frac{d\log{P}}{dr},$$

\noindent and converting to non-logarithmic derivatives, this becomes

\begin{equation}
    N^2 = -\frac{g}{\rho}\parderiv{\rho}{S}{P}{Y}\bigg[\fullderiv{S}{r} - \parderiv{S}{Y}{\rho}{P}\fullderiv{Y}{r} \bigg],
\end{equation}

\noindent which is the expression found in \cite{Lattimer1981}. Using the relation

\begin{equation}
    \parderiv{\rho}{S}{P}{Y} = -\frac{\chi_T}{\chi_{\rho}}\frac{\rho}{c_P},
\end{equation}

\noindent the Brunt V\"{a}is\"{a}l\"{a} frequency (squared) can now be written as 

\begin{equation}
    N^2 = \frac{g}{c_P}\frac{\chi_T}{\chi_{\rho}}\bigg[\fullderiv{S}{r} - \parderiv{S}{Y}{\rho}{P}\fullderiv{Y}{r} \bigg]\, .
\end{equation}

\bibliography{references_free.bib}

\end{document}